\documentclass[sigconf]{acmart}

\makeatletter
\def\@ACM@checkaffil{
    \if@ACM@instpresent\else
    \ClassWarningNoLine{\@classname}{No institution present for an affiliation}%
    \fi
    \if@ACM@citypresent\else
    \ClassWarningNoLine{\@classname}{No city present for an affiliation}%
    \fi
    \if@ACM@countrypresent\else
        \ClassWarningNoLine{\@classname}{No country present for an affiliation}%
    \fi
}
\makeatother

\renewcommand\footnotetextcopyrightpermission[1]{} 

\AtBeginDocument{%
  }

\usepackage{graphicx}
\usepackage{float}
\usepackage{subfigure}
\usepackage{multirow}
\usepackage[ruled,vlined,linesnumbered]{algorithm2e}

\begin{document}

\title{Binary Embedding-based Retrieval at Tencent}


\author{Yukang Gan}
\authornote{The authors contributed equally to this work.}
\email{brucegan@tencent.com}
\affiliation{%
  \institution{ARC Lab, Tencent PCG}
}

\author{Yixiao Ge}
\authornotemark[1]
\email{yixiaoge@tencent.com}
\affiliation{%
  \institution{ARC Lab, Tencent PCG}
}

\author{Chang Zhou}
\authornotemark[1]
\email{chanzhou@tencent.com}
\affiliation{%
  \institution{Tencent Video, PCG}
}

\author{Shupeng Su}
\email{pennsu@tencent.com}
\affiliation{%
  \institution{ARC Lab, Tencent PCG}
}

\author{Zhouchuan Xu}
\email{atuerxu@tencent.com}
\affiliation{%
  \institution{Tencent Search, PCG}
}

\author{Xuyuan Xu}
\email{evanxyxu@tencent.com}
\affiliation{%
  \institution{Tencent Video, PCG}
}

\author{Quanchao Hui}
\email{andrewshui@tencent.com}
\affiliation{%
  \institution{Tencent Search, PCG}
}

\author{Xiang Chen}
\email{joshuaxchen@tencent.com}
\affiliation{%
  \institution{Tencent Search, PCG}
}

\author{Yexin Wang}
\email{yexinwang@tencent.com}
\affiliation{%
  \institution{Tencent Video, PCG}
}

\author{Ying Shan}
\email{yingsshan@tencent.com}
\affiliation{%
  \institution{ARC Lab, Tencent PCG}
  \institution{Tencent Search, PCG}
}





\renewcommand{\shortauthors}{Yukang Gan and Yixiao Ge, et al.}

\begin{abstract}
Large-scale embedding-based retrieval (EBR) is 
the cornerstone of search-related industrial applications. 
Given a user query, the system of EBR aims to identify relevant information from a large corpus of documents that may be tens or hundreds of billions in size.
The storage and computation turn out to be expensive and inefficient with massive documents and high concurrent queries, making it difficult to further scale up.
%
%

To tackle the challenge, we propose a binary embedding-based retrieval (BEBR) engine equipped with a recurrent binarization algorithm that enables customized bits per dimension.
Specifically, we compress the full-precision query and document embeddings, formulated as float vectors in general, into a composition of multiple binary vectors using a lightweight transformation model with residual multilayer perception (MLP) blocks.
The bits of transformed binary vectors are jointly determined by the output dimension of MLP blocks (termed $m$) and the number of residual blocks (termed $u$), \textit{i.e.}, $m\times(u+1)$.
We can therefore tailor the number of bits for different applications to trade off accuracy loss and cost savings.
Importantly, we enable task-agnostic efficient training of the binarization model using a new embedding-to-embedding strategy, \textit{e.g.}, only 2 V100 GPU hours are required by millions of vectors for training. We also exploit the compatible training of binary embeddings so that the BEBR engine can support indexing among multiple embedding versions within a unified system. To further realize efficient search, we propose Symmetric Distance Calculation (SDC) to achieve lower response time than Hamming codes. The technique exploits Single Instruction Multiple Data (SIMD) units widely available in current CPUs. 

We successfully employed the introduced BEBR to web search and copyright detection of Tencent products, including Sogou, Tencent Video, QQ World, \textit{etc}. The binarization algorithm can be seamlessly generalized to various tasks with multiple modalities, for instance, natural language processing (NLP) and computer vision (CV). Extensive experiments on offline benchmarks and online A/B tests demonstrate the efficiency and effectiveness of our method, significantly saving $30\%\sim50\%$ index costs with almost no loss of accuracy at the system level\footnote{Code is publicly available at \url{https://github.com/ganyk/BEBR}.}. 


\end{abstract}





\acmConference[KDD'23]{ACM SIGKDD Conference on Knowledge
Discovery and Data Mining}{August 6--10, 2023}{Long Beach, California, USA}

\acmDOI{}
\acmISBN{}
\acmPrice{}

\maketitle
\pagestyle{plain}

\section{Introduction}
\begin{figure}
\centering
\includegraphics[width=0.45\textwidth]{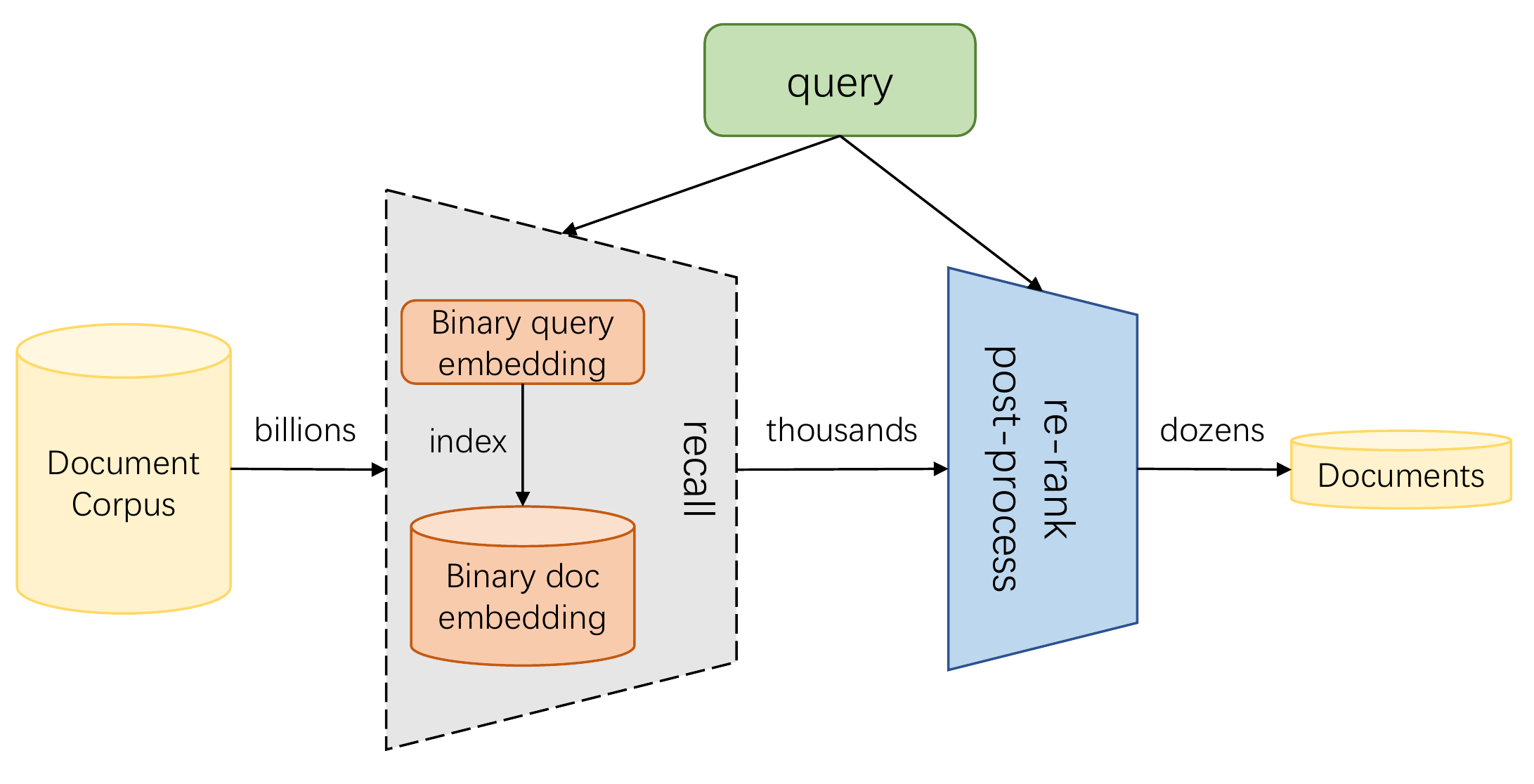} 
\vspace{-5pt}
\caption{A brief structure of the Tencent search system, composed of the introduced binary embedding-based retrieval (BEBR) engine for recall and a re-rank post-process layer.
}\vspace{-5pt}
\label{Fig.overview}
\end{figure}

With the development of deep learning, embedding-based retrieval (EBR) achieves great advances in real-world applications, such as web search \cite{wu2020zero}, social search~\cite{huang2020embedding}, e-commerce search~\cite{li2021embedding}, \textit{etc}. 
Generally speaking, a typical industrial search-related system is composed of a ``recall-rerank'' architecture (as demonstrated in Figure \ref{Fig.overview}), in which the efficiency of the recall module with EBR algorithms is the bottleneck of the whole system as it needs to process massive documents.
Unlike conventional inverted index-based term matching~\cite{robertson2009probabilistic} that measures similarity through lexical analysis, EBR represents queries and documents as dense feature vectors. 
Given a query, EBR retrieves a set of relevant documents according to their embedding similarities in the latent space.
The enormous scale of documents and high concurrent queries pose great challenges to an industrial EBR system, including retrieval latency, computation cost, storage consumption, and embedding upgrades.

There are previous attempts to develop more efficient EBR systems with advanced ANN (Approximate Nearest Neighbor) algorithms, \textit{e.g.}, HNSW \cite{malkov2018efficient}. 
Though the achievements in saving computations, they need elaborate designs to be adapted and plugged into existing systems.
Given the large number and variety of EBR systems for Tencent products, the development costs of upgrading all the existing ANN algorithms are non-negligible and even unaffordable.
Toward this end, we focus on the most fundamental component of EBR, that is, embedding, also known as representation learning in the deep learning community.
Properly compressed embeddings are compatible with mainstream ANN algorithms and can be seamlessly integrated into existing EBR systems.

In this work, we propose a binary embedding-based retrieval (BEBR) engine that has several appealing benefits: 
(i) customizable embedding compression rates to receive a trade-off between accuracy and costs;
(ii) a task-agnostic and modal-agnostic efficient training paradigm for easy generalization and data security protection;
(iii) a free embedding upgrading mechanism with backward compatibility, \textit{i.e.}, no need to refresh the index.
BEBR has been well deployed on multiple Tencent products equipped with various ANN algorithms (\textit{e.g.}, IVF \cite{moffat1996self}, HNSW \cite{malkov2018efficient}) with almost no accuracy loss and 30$\sim$50\% cost savings at the system level.

Specifically, inspired by recurrent binary embeddings~\cite{shan2018recurrent} that progressively refine a base binary vector with binary residual vectors to meet task accuracy, BEBR develops a universal binarization algorithm with state-of-the-art performances across modalities. Rather than the simple linear transformations used in \cite{shan2018recurrent}, BEBR adopts multilayer perception (MLP) blocks with non-linear layers (\textit{i.e.}, ReLU \cite{agarap2018deep}) for both binarization (float$\to$binary) and reconstruction (binary$\to$float) in the recurrent learning paradigm.
As illustrated in Figure \ref{Fig.RBE}, the binarization, reconstruction, and residual blocks together form the recurrent binarization module with a customized number of loops, \textit{i.e.}, bits per dimension.
The recurrent binary embeddings with richer representations are much more discriminative than ordinary hash vectors \cite{cao2017hashnet}.


In previous works, the binarization (or hashing) module is usually optimized end-to-end with the backbone network, \textit{e.g.}, CNNs \cite{su2018greedy} for vision, and Transformers \cite{ou2021refining} for text. The training is expensive considering the heavy backbone models for accurate retrieval. 
We, therefore, introduce an efficient training paradigm that requires only floating-point vectors as input.
The lightweight binarization module is trained individually without accessing the backbone models, forming a universal training procedure for all the modalities and tasks.
To enable effective representation learning in such an embedding-to-embedding paradigm, we use contrastive learning with queue-based hard negative mining as the training objectives.



Besides the index costs, large-scale EBR systems heavily suffer from the computational overhead required by embedding model upgrades. In particular, all the embeddings in the index need to be re-extracted before the deployment of a new model, which may take weeks or even months for industrial applications. Thanks to the pioneering work in compatible representation learning \cite{shen2020towards,hu2022learning}, we take the first step to investigate compatible training of binary embeddings. Equipped with backward-compatible learning, our BEBR engine is able to harvest the benefit of the new model immediately, \textit{i.e.}, the queries encoded by the new model can be directly indexed among the old index.


We further propose Symmetric Distance Calculation (SDC) of recurrent binary embeddings, a novel technique that achieves significant speedup over 
the conventional Hamming-based distance calculation
in \cite{shan2018recurrent}.
SDC leverages the in-register cache to perform fast SIMD (Single Instruction Multiple Data) look-up instructions and is especially in favor of CPU-based computation platforms. 
Comprehensive experiments on public benchmarks, internal datasets, and online A/B tests on Tencent products fully demonstrate the effectiveness of our BEBR engine. It has been successfully deployed on almost all ranges of index-based applications in Tencent PCG, including web search (Sogou), video search, copyright detection, video recommendation, \textit{etc}.


The contributions are four-fold.
\begin{itemize}
    \item[$\bullet$] We propose a binary embedding-based retrieval (BEBR) engine that efficiently indexes among tens of billions of documents in Tencent products. The proposed method can be equipped with various ANN algorithms and integrated into existing systems seamlessly.
    \item[$\bullet$] BEBR drastically reduces both the memory and disk consumption while achieving superior retrieval performance with the benefit of tailored recurrent binarization and symmetric distance calculation.
    \item[$\bullet$] BEBR develops a universal training paradigm for all modalities without accessing the raw data and backbone networks, \textit{i.e.}, the binary embeddings are trained efficiently in a task-agnostic embedding-to-embedding manner.
    \item[$\bullet$] BEBR enables backward-compatible upgrades of embedding models, that is, the new model can be immediately deployed without refreshing the index embeddings. We for the first time study compatible learning on binary embeddings.
\end{itemize}


\begin{figure*}
\centering
\includegraphics[width=0.75\textwidth]{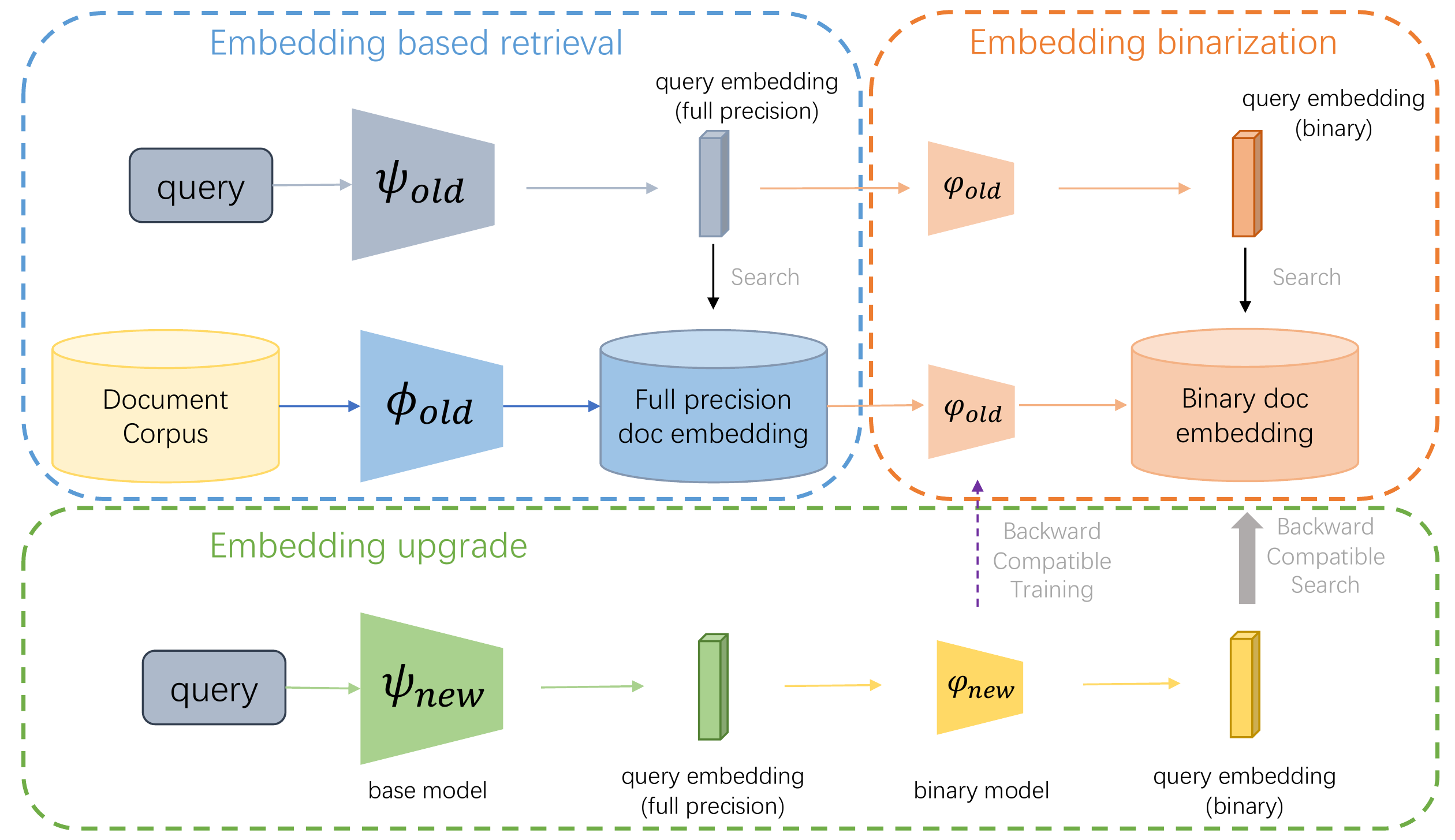} 
\caption{Our binary embedding-based retrieval (BEBR) framework. The full-precision float embeddings, extracted by the backbone networks, are transformed to recurrent binary vectors using a parametric binarization module $\varphi$ in a task-agnostic embedding-to-embedding manner. BEBR enables backfill-free upgrades for the binarization model, that is, the new model can be immediately deployed for encoding better query embeddings without refreshing the index. 
}
\label{Fig.framework}
\end{figure*}

\section{Related work}
\subsection{Embedding-based Retrieval in Search}
Representation learning for embedding-based retrieval has attracted much attention in academia and industry given its great success in various domains.
For example, as a social search engine, Facebook learns semantic embeddings for personalized search, which serves an EBR system with ANN parameter tuning \cite{huang2020embedding}. 
For e-commerce search, Taobao proposes a Multi-Grained Deep Semantic Product Retrieval (MGDSPR) \cite{li2021embedding} system to capture the relation between user query semantics and his/her personalized behaviors. JD proposes Deep Personalized and Semantic Retrieval (DPSR)\cite{zhang2020towards} to combine text semantics and user behaviors. Amazon develops a Siamese network to address the semantic gap problem for semantic product retrieval \cite{nigam2019semantic}. For web search, Google adopts zero-shot heterogeneous transfer learning in the recommendation system to improve search performance \cite{wu2020zero}.
While none of the aforementioned methods studies the trade-off between performance and costs in EBR implementation, this paper discusses the binary embedding-based retrieval system, which can achieve near-lossless performance with significant cost reduction.

\subsection{ANN Methods}
Plenty of research has been devoted to developing efficient ANN algorithms. 
Some of them build graphs from datasets to avoid the exhaustive search where each vertex in the graph is associated with a data point. Others encode the embeddings into compact codes to reduce memory consumption and speed up distance calculation.
 
Specifically, the graph-based methods generally leverage the k-Nearest Neighbor graph to allow fast navigation in the index. For example, ~\cite{malkov2012scalable} proposes a proximity graph algorithm called Navigable Small World (NSW), which utilizes navigable graphs. Hierarchical NSW ~\cite{malkov2018efficient} offers a much better logarithmic complexity scaling with a controllable hierarchy. More recently, Navigating Spreading-out Graph (NSG) ~\cite{fu2017fast} proposes a novel graph structure that guarantees very low search complexity. It strongly outperforms previous state-of-the-art approaches. Although these graph-based algorithms achieves high search performance at high precision, they need more memory space and data-preprocessing time than product quantization and hashing-based methods. Therefore, in frequent updating scenarios, building the index from the graph-based algorithm on the large dataset is impractical.

Product quantization (PQ) ~\cite{jegou2010product} decomposes the space into a Cartesian product of low dimensional subspaces and quantizes each subspace separately. Cartesian K-means (CKM)~\cite{norouzi2013cartesian} and Optimized Product Quantizers (OPQ)~\cite{ge2013optimized} extends the idea of space decomposition and optimizes the sub-space decomposition by arbitrarily rotating and permutating vector components. In addition, variants of quantization models ~\cite{babenko2014additive,babenko2015tree,zhang2014composite} inspired by PQ have been proposed. These models offer a lower quantization error than PQ or OPQ. Recently, PQ Fast Scan~\cite{andre2019derived} pioneers the use of SIMD for Asymmetrical Distance Calculation (ADC) evaluation, and later works \cite{andre2019quicker} \cite{blalock2017bolt} \cite{andre2017accelerated} are proposed to optimize the quantization scheme for achieving lower search latency for indexed databases. Inspired by the ADC techniques, we propose a symmetrical distance calculation (SDC) to enable efficient search in BEBR retrieval.

Hashing-based algorithms have recently gained popularity due to their advantages in computing and storage. 
Existing hashing-based algorithms can be organized into two categories: locality-sensitive hashing ~\cite{charikar2002similarity, indyk1998approximate} and learning to hash. Locality-sensitive hashing (LSH) maps the original data into several hash buckets where similar items are more likely to fall into the same bucket. Despite the fact that tremendous efforts~\cite{indyk1998approximate, broder1997syntactic, gan2012locality, ji2012super, li2006very} have been exerted to improve the performance of LSH. It still requires multiple hash tables to maintain the recall rate of search, which limits its application on large-scale datasets. Learning to hash is a data-dependent approach that learns hash functions from a specific dataset. With the development of deep learning, many methods ~\cite{cao2017hashnet, fan2020deep, li2015feature, liu2018deep, su2018greedy} adopt the powerful capabilities of deep neural network (DNN) to learn the complex hash functions and obtain binary codes through an end-to-end manner. Instead of converting data into normal binary vectors, ~\cite{shan2018recurrent} proposes recurrent binary embedding to achieve a balanced goal of retrieval performance, speed, and memory requirement. Specifically, it progressively adds a residual binary vector to the base binary vector. A GPU-based k-nearest neighbor (K-NN) selection algorithm is also implemented, enabling exhaustive real-time search on billion-scale data sets. In this paper, towards the goal of efficient and low-cost embedding quantization, we use the off-the-shelf float-point-based embeddings as input to learn recurrent binary embeddings. Furthermore, we provide an efficient method to calculate the distance between recurrent binary embeddings using CPUs, the most common computing devices in the industrial retrieval system.

\subsection{Compatibility of Deep Neural Networks}

Compatible representation learning aims at making embeddings comparable across models. 
It has attracted increasingly extensive attention in industry and academia due to its ability to reduce computation costs in embedding upgrades. Specifically, there are two types of compatibility: cross-model and backward compatibility. Cross-model compatibility learning usually trains transformation modules to map the embeddings from different models into a common space. $R^3AN$ ~\cite{chen2019r3} firstly introduces the problem of cross-model compatibility in face recognition and tackles it by learning a transformation that transforms source features into target features through a process of reconstruction, representation, and regression. ~\cite{wang2020unified} strikingly ameliorates the cross-model compatibility performance by coming up with a unified representation learning framework. In detail, they design a lightweight Residual Bottleneck Transformation (RBT) module and optimize it with a classification loss, a similarity loss, and a KL-divergence loss. 

While cross-model compatibility handles embeddings from different models, backward compatibility fixes its attention on model updates where new models are trained with additional compatibility constraints. In a sense, it enables compatibility between new embeddings and old embeddings without any extra transformation processes. ~\cite{shen2020towards} is the first work that utilizes backward compatibility to conduct model upgrades. It introduces an influence loss when training the new model to enable direct comparison between new embeddings and old embeddings. Under this backward compatibility framework, several works make attempts to ameliorate the performance of backward compatibility by leveraging hot-refresh backward-compatible model upgrades ~\cite{zhang2022hot}, asymmetric retrieval constraints ~\cite{budnik2021asymmetric}, embedding cluster alignment loss ~\cite{meng2021learning}, and neural architecture search ~\cite{duggal2021compatibility}. In this paper, we adopt backward compatibility training to learn backward compatible binary embeddings. To our best knowledge, this is the first time compatible learning has been applied to binary embedding.

\section{Binary Embedding-based Retrieval}

\subsection{Preliminary}

Given a query $q$ (generally a text in a web search or a video in copyright detection), an embedding-based retrieval (EBR) system aims to rank the documents $\left \{ d_0, d_1,\cdots,d_n \right \}$ according to their similarities. 
There are two key factors in EBR, the embedding model(s) and the distance calculation metric $\mathcal{D}(\cdot, \cdot)$. 
The cosine similarity is widely used as $\mathcal{D}(\cdot, \cdot)$ for full precision embeddings (float vectors).
Formally, the similarity between a certain query and a document is
\begin{equation}
\mathcal{S}_\text{EBR}(q, d_k) = \mathcal{D}\left(\psi(q), \phi(d_k)\right), ~~~\forall k\in\{1,\cdots,n\},
\end{equation}
where $\psi$ and $\phi$ are embedding models for queries and documents. $\psi$ and $\phi$ can be designed to be identical or distinct to handle different retrieval tasks with homogeneous or heterogeneous input~\cite{guo2022semantic}.
For example, we may have ResNet~\cite{he2016deep} for image data and Transformer~\cite{devlin2018bert} for text. 
Without loss of generality, we consider homogeneous architecture (\textit{i.e.}, $\psi$ and $\phi$ are identical and both denoted as $\phi$) in the following cases.
To tackle the billion-level indexing at a moderate cost, we introduce binary embedding-based retrieval (BEBR) engine with much more efficient similarity calculation between queries and documents, such as
\begin{equation}\label{eq:sim_bebr}
\mathcal{S}_\text{BEBR}(q, d_k) = \mathcal{D}\left(\phi\circ\varphi(q), \phi\circ\varphi(d_k)\right), ~~~\forall k\in\{1,\cdots,n\},
\end{equation}
where $\varphi(\cdot)$ is the binarization process and is generally realized by a parametric network.
In the following sections, we will introduce the detailed designs of $\varphi(\cdot)$ in Section \ref{sec:rb} and $\mathcal{D}(\cdot, \cdot)$ in Section \ref{sec:deploy}.

\begin{figure}
\centering
\includegraphics[width=0.45\textwidth]{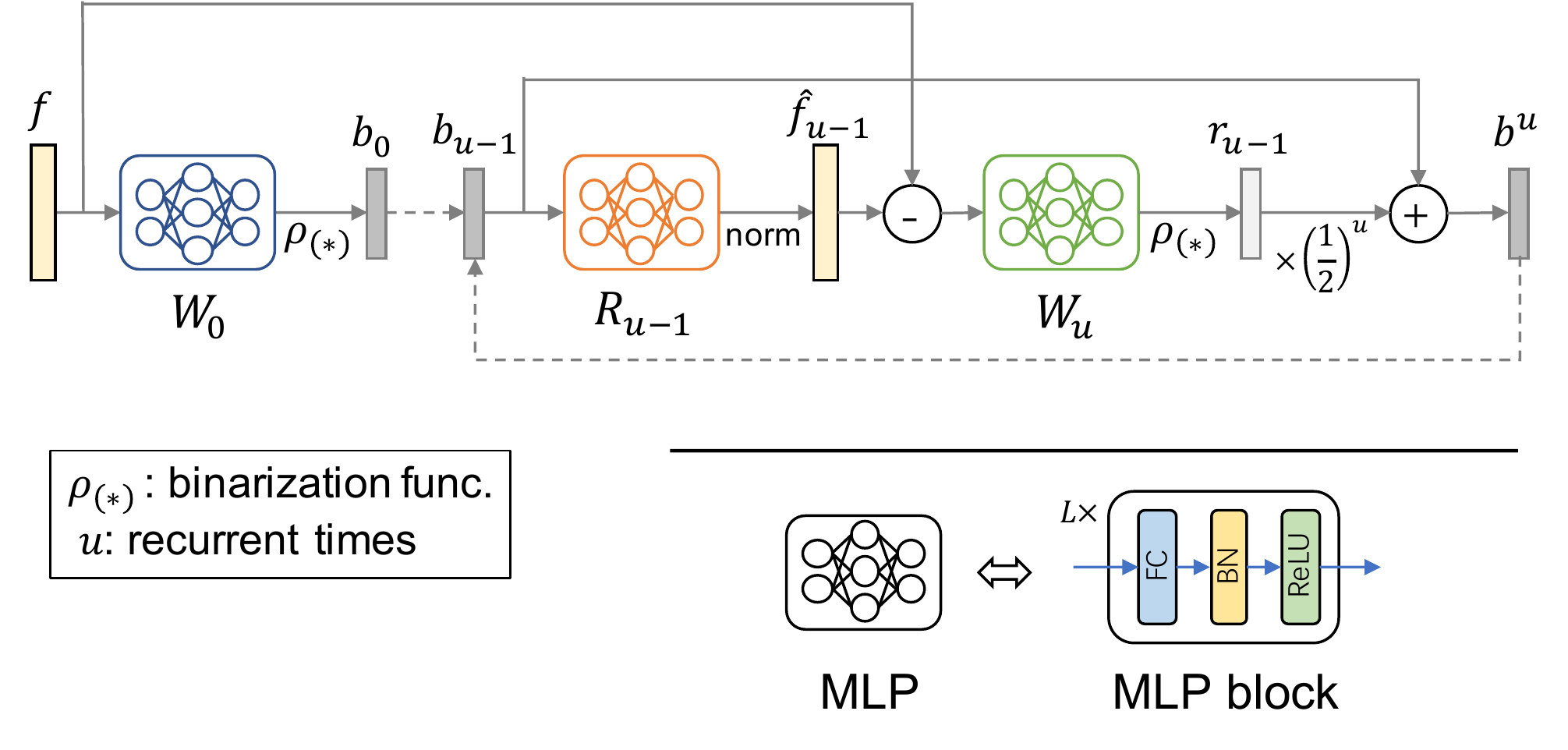} 
\vspace{-10pt}
\caption{The architecture of recurrent binary embedding model $\varphi$. $\ominus$ and $\oplus$ denote minus and plus operations between two input embeddings, respectively. }\vspace{-5pt}
\label{Fig.RBE}
\end{figure}

\subsection{Recurrent Binarization}\label{sec:rb}

\subsubsection{Architecture}

To tackle the problem of learning to binarization, the straightforward solution is to adopt hashing networks \cite{zheng2020deep} ended up with a binarization function $\rho$, which plays an important role in converting float vectors into binary ones composed of either $-1$ or $+1$. In the forward pass, $\rho$ is formulated as
$$ \rho(x)\equiv\text{sign}(x)=\left\{
\begin{aligned}
-1 &, ~~~ x\le0 \\
1 &, ~~~ x>0
\end{aligned}
\right.
$$
Since the gradient of the sign function vanishes and thus cannot be back-propagated directly, \cite{courbariaux2016binarized} introduced a straight-through estimator (STE) that takes the gradient of the identity function instead, that is, $\rho'(x)=1$ when $|x|\le1$ and $\rho'(x)=0$ otherwise.
Conventional hashing methods generally convert the float vectors to binary embeddings once with some learnable layers and a binarization function introduced above. However, such methods suffer from unsatisfactory performance due to the limited representation ability of the hash codes that only have $-1$ or $+1$ values, \textit{i.e.}, 1 bit per dimension.
Thanks to the pioneering work \cite{shan2018recurrent}, binary embeddings are able to be progressively refined with customized bits per dimension.

Specifically, following the insight of using residual operations to gradually narrow the gap between original float vectors and the learned binary ones, we introduce a recurrent binarization module with customized loops, as demonstrated in Figure \ref{Fig.RBE}.
There are three main components, including binarization block, reconstruction block, and residual block. The binarization block performs almost the same as conventional hashing networks, where a binary embedding $b_0$ is encoded from the float vector $f$ as input: $b_0 = \rho(\boldsymbol{W}_0(f))\in \{-1, 1\}^{m}$,
where $\rho$ is the binarization function, and $\boldsymbol{W}_0$ is the multi-layer perception (MLP) that consists of linear, batch normalization, and ReLU layers.
The encoded binary embedding $b_0$ is then reconstructed back to float vectors, such as $\hat{f}_0 = \|\boldsymbol{R}_0(b_0)\|$,
where $\boldsymbol{R}_0$ is also multi-layer perception (MLP).
The residual between the original $f$ and reconstructed $\hat{f}_0$, therefore, reflects the representation loss of the binarization process, which can be further narrowed by repeating the above steps to binarize the residual parts.
The residual binary vector can be formulated as $r_0=\rho(\boldsymbol{W}_1(f-\hat{f}_0))$, which is further added to the base binary vector $b_0$ via $b_1=b_0+\frac{1}{2}r_0$. The weight $\frac{1}{2}$ is chosen to ease the similarity calculation with only $xor$ and $popcount$ (find detailed derivation from the original paper of \cite{shan2018recurrent}).

Until now, we have introduced the process of recurrent binarization when the loop is set as $1$, \textit{i.e.}, repeat once. In real-world applications, the loop can be customized to trade off accuracy and efficiency. Formally, the whole process of recurrent binarization with $u \geq 1$ loops can be defined as
\begin{equation}
\text{base binarization:~~~}b_0 = \rho(\boldsymbol{W}_0(f)), \nonumber
\end{equation}
\begin{equation}
\text{residual binarization loops:~~~}\left\{
\begin{aligned}
\hat{f}_{u-1} &= \|\boldsymbol{R}_{u-1}(b_{u-1})\|, \\
r_{u-1} &= \rho(\boldsymbol{W}_{u}(f - \hat{f}_{u-1})), \\
b_u &= b_{u-1} + 2^{-u} \boldsymbol{r}_{u-1}.  \nonumber
\end{aligned}
\right.
\end{equation}
The recurrent binary embedding $b_u$ is the output of the binarization process in Eq. (\ref{eq:sim_bebr}), \textit{i.e.}, $b_u=\varphi(f)$ given $f=\phi(q)$ or $f=\phi(d_k)$. Given the output dimension of $\boldsymbol{W}$ as $m$, the overall bits of $b_u$ can be calculated as $m\times(u+1)$.


\subsubsection{Task-agnostic training}


As shown in Eq. (\ref{eq:sim_bebr}), the backbone network $\phi$ and the binarization module $\varphi$ are commonly jointly optimized in an end-to-end manner in previous learning to hash methods \cite{fan2020deep}. Though feasible, the training is not efficient given the heavy backbone network for accurate representation learning, and task-dependent as the raw data (\textit{e.g.}, text, images) must be accessed for training end-to-end, rendering an inflexible solution, especially for the data-sensitive applications.
To tackle the challenge, we introduce a universal training solution that only requires the float vectors as input, \textit{i.e.}, extracting the embeddings using off-the-shelf backbone networks $\phi$. The binarization module $\varphi$ is therefore trained alone in a task-agnostic and modality-agnostic manner. The objective function of such embedding-to-embedding training can be formulated as
\begin{align}
    \mathop{\arg\min}_{\varphi} \ \ \mathcal{L}(\mathcal{F}; \varphi),
\end{align}
where $\mathcal{F}$ is the set of all float vectors for training.

Given the great success of contrastive loss \cite{chen2020simple} in representation learning research, we adopt an NCE-form contrastive objective to regularize the binarization module,
\begin{align}\label{Eq.binary_loss}
    \mathcal{L}(\mathcal{F}; \varphi)=\frac{1}{|\mathcal{F}|}\sum_{f\in\mathcal{F}}-\log\frac{\exp\langle\varphi(f),\varphi(k_+)\rangle}{\sum_{k\in\mathcal{B}}\exp\langle\varphi(f),\varphi(k)\rangle},
\end{align}
where $k$ is the float features within the same batch $\mathcal{B}$ as the anchor $f$. $k_+$ is the positive sample constructed from another augmented view of an image or the query-document pair from the web. $\langle\cdot,\cdot\rangle$ is cosine similarity between recurrent binary embeddings.
Besides the positive pairs collected by manual annotations or user behaviors, we employ hard negative mining to further improve the discriminativeness of the learned binary embeddings.

 Hard negative mining has proven to be a useful technique for improving the accuracy of classification tasks~\cite{shrivastava2016training, devlin2018bert, schroff2015facenet} in the deep learning community. 
 Recent work~\cite{huang2020embedding, li2021embedding} on semantic retrieval has also successfully applied this technique to improve the retrieval accuracies. 
 There are online and offline hard negative mining approaches
 to collect hard enough negative samples and improve the model's ability to identify similar but irrelevant query-document pairs. 
 Online hard negative mining is efficient as it is conducted on the fly within mini-batches. Offline hard mining is performed off-the-shelf before each training epoch and is extremely time-consuming, even with the help of the ANN algorithm. However, offline mining is proven to be more effective as it can search among the whole training set and discover the most difficult samples.
 How to enable global hard mining as the offline method while at the same time maintaining the efficiency of online methods turn out to be a challenging but critical problem.

Inspired by \citet{he2020momentum}, we tackle this problem by maintaining a queue $\mathcal{Q}\in\mathbb{R}^{L\times m}$ of negative sample embeddings. Specifically, we extend the mini-batches with a fix-length (\textit{i.e.}, $L$) queue (about 16$\times$ larger than the mini-batch) and mine hard samples in the queue on the fly. 
At each training step, the binary embeddings of the current mini-batch are added to the queue, and the oldest mini-batch in the queue is removed if the maximum capacity is reached. Note that we perform momentum updates of the binarization module to encode embeddings for the queue in order to keep latent consistency among different batches, following the practice in \cite{he2020momentum}.
We select the top-$k$ hardest negative samples in the queue for contrastive objectives in Eq. (\ref{Eq.binary_loss}), \textit{i.e.}, the samples that receive the highest similarity scores with the anchor feature. 
Therefore, the set of training samples $\mathcal{B}$ in Eq. (\ref{Eq.binary_loss}) becomes
\begin{equation}
\mathcal{B} = \{ k_+, \kappa(\mathcal{Q})\},
\end{equation}
where $\kappa(\mathcal{Q})$ denotes the operation for selecting top-$k$ hardest negative samples from $\mathcal{Q}$.


Once $\varphi$ is learned, the recurrent binary embeddings for queries and documents can be produced in an efficient embedding-to-embedding paradigm. Both training and deployment processes are task-agnostic since only full-precision embeddings are needed as input, which enables universal embedding binarization across all the modalities and tasks.

\subsubsection{Backward-compatible training}
As illustrated in Figure \ref{Fig.overview}, 
the retrieval stage needs to process billions or trillions of documents. The huge scale of data poses challenges in embedding upgrades since all the index embeddings need to be re-extracted before the deployment of a new model. Such a process is quite time-consuming and computationally expensive.
In this paper, we for the first time investigate the potential of backward-compatible learning \cite{shen2020towards} with binary embeddings. To be specific, compatible learning requires the embeddings encoded by the old model and the new model to be interchangeable in a consistent latent space. Embedding model upgrades with backward compatibility can deploy the new model immediately without refreshing the index, that is, the new query embeddings can be directly compared with the old document embeddings. The upgrading objective can be formulated as
\begin{align}
\mathcal{S}_\text{BEBR-BC}(q_\text{new}, d^+_\text{old}) & \geq \mathcal{S}_\text{BEBR}(q_\text{old}, d^+_\text{old}), \\
\mathcal{S}_\text{BEBR-BC}(q_\text{new}, d^-_\text{old}) & \leq \mathcal{S}_\text{BEBR}(q_\text{old}, d^-_\text{old}),
\end{align}
where $d^+$ denotes relevant documents to user query $q$, and $d^-$ denotes the irrelevant ones. $\mathcal{S}_\text{BEBR-BC}(\cdot, \cdot)$ calculates the similarity between the new binary embedding of query and old binary embedding of the document, which is formulated as:
\begin{equation}
\begin{split}
\mathcal{S}_\text{BEBR-BC}(q, d_k) = \mathcal{D}\left(\tilde{\phi}\circ\varphi_{\text{new}}(q), \phi\circ\varphi_{\text{old}}(d_k)\right)& \\ 
\forall k\in\{1,\cdots,n\}&,
\end{split}
\end{equation}
where $\varphi_{\text{new}}(\cdot)$ is the new version of the recurrent binary transformation module and $\varphi_{\text{old}}(\cdot)$ is the old one, $\tilde{\phi}$ denotes a new or identical float backbone model determined by specific applications. BEBR-BC stands for backward compatible BEBR system.


The training objective can be formulated as
\begin{align}
    \mathop{\arg\min}_{\varphi_\text{new}} \ \ \mathcal{L}(\mathcal{F}; \varphi_\text{new})+\mathcal{L}_\text{BC}(\mathcal{F}; \varphi_\text{new}, \varphi_\text{old}),
\end{align}
where $\mathcal{L}$ is the same as Eq. (\ref{Eq.binary_loss}), and
$\mathcal{L}_\text{BC}$ is also in the form of an NCE loss but across old and new models, \textit{i.e.},
\begin{align}\label{Eq.binary_bc_loss}
    \mathcal{L}_\text{BC}(\mathcal{F};& ~~\varphi_\text{new}, \varphi_\text{old})\nonumber\\
    &=\frac{1}{|\mathcal{F}|}\sum_{f\in\mathcal{F}}-\log\frac{\exp\langle\varphi_\text{new}(\tilde{f}),{\varphi}_\text{old}(k_+)\rangle}{\sum_{k\in\mathcal{B}}\exp\langle\varphi_\text{new}(\tilde{f}),{\varphi}_\text{old}(k)\rangle}.
\end{align}
$\tilde{f}$ is encoded by $\tilde{\phi}(\cdot)$. $\varphi_\text{new}$ is optimized individually with the other parametric modules fixed.
$\mathcal{L}$ maintains self-discrimination of the new binarization model while $\mathcal{L}_\text{BC}$ regularizes the cross-model compatibility.
Queue-based hard mining is also applied for $\mathcal{L}_\text{BC}$.




\subsection{Deployment}\label{sec:deploy}

\subsubsection{Dot product of recurrent binary embeddings: A revisit} \label{SSec.binary_and_quantization}
In \citet{shan2018recurrent}, the cos similarity of recurrent binary embedding is decomposed into the dot product of hash codes as in Eq. (\ref{eq18}), where the subscript $q$ and $d$ denote the query and documentation. Thus, the calculation of hash codes can be implemented efficiently with the bit-wise operation as in Eq. (\ref{eq19}), where ${x}$, ${y}$, are binary vectors in $\{1,-1\}^m$, $popc$, $\wedge$ and $>>$ are the population count, XOR, and logical right shift operations. 
 \begin{equation} \label{eq18}
 \begin{aligned}
\mathcal{D}(b^q_u,b^d_u)\propto &\frac{1}{||b^d||}(b^q_0\cdot b^d_0 + 
\sum_{j=0}^{u-1}\sum_{i=0}^{u-1} (\frac{1}{2})^{j+i+2} r^q_j\cdot r^d_i \\
& + \sum_{j=0}^{u-1}(\frac{1}{2})^{j+1}b^q_0\cdot r^d_j+\sum_{i=0}^{u-1}(\frac{1}{2})^{i+1}b^d_0\cdot r^q_i)
\end{aligned}
\end{equation}
\begin{equation} \label{eq19}
x\cdot y = (popc(x\wedge y) >> 1) + m
\end{equation}
Although the bit-wise operation is fast with population count, the computation complexity grows rapidly with the increase of ${u}$. Hence, it relies on GPU to offer high performance, and an optimized k-NN selection algorithm is developed. 

\subsubsection{Symmetric distance calculation (SDC)}
Unfortunately, the GPU-Enabled NN search algorithm limits its usefulness and applicability in practical cases. In this paper, we develop a Symmetric Distance Calculation (SDC) of recurrent binary embedding around the CPU platform, which is applicable to most scenarios. Specifically, SDC allows computing the distance between the uncompressed recurrent binary features. It relies on SIMD in-register shuffle operation to provide a high-performance calculation procedure, which can be combined with inverted indexes. For simplicity's sake, we explain the SDC uses 128-bit SIMD in the following content.

 Similar to \cite{andre2017accelerated},\cite{andre2019quicker}, and \cite{blalock2017bolt}, SIMD registers and in-register shuffles are used to store lookup tables and perform lookups. However, these methods use sub-quantizers to obtain different centroids without normalization during calculation. Therefore, algorithmic changes are  required to obtain the fixed centroids and magnitude of embeddings for normalization. More specifically, SDC relies on 4-bit code as a basic unit and uses 8-bit integers for storing lookup tables. The resulting lookup tables comprise 16 8-bits integers (128 bits). Once lookup tables are stored in SIMD registers, in-register shuffles can perform 16 lookups in 1 cycle, enabling large performance gains. 
 
\begin{figure}[t]
\centering
\includegraphics[width=0.45\textwidth]{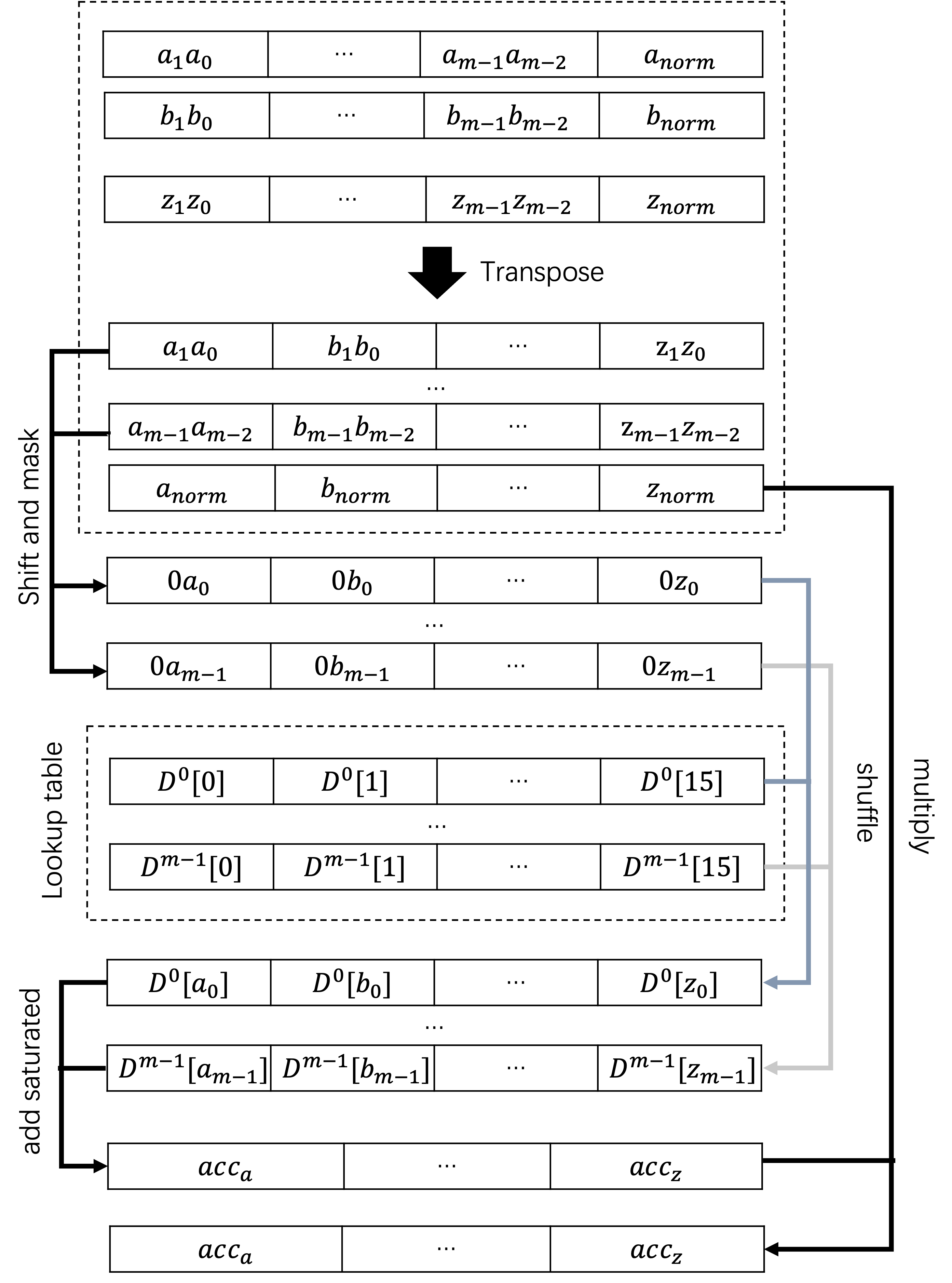} 
\vspace{-5pt}
\caption{Symmetric distance calculation (SDC) using SIMD calculation.} \vspace{-5pt}
\label{figure_rbe_adc_calculate}
\end{figure}

 
\textbf{Memory layout.} By setting $u\in\{2,4\}$, we first generate the recurrent binary vectors and organize an inverted list of features with the standard memory layout. As shown in the upper of Figure \ref{figure_rbe_adc_calculate}. $a_i$ is the 4-bit code, and $a_{norm}$ is the quantized magnitude value of the vector appended at the end. Notably, for $u=2$, the $a_i$ represents two adjacent dimensions of the feature. To efficiently shuffle lookup tables, the standard memory layout of inverted lists needs to be transposed because the transposed data are contiguous in memory, and the SIMD register can be loaded in a single memory read. This transition process is performed offline and does not influence the search speed. 
 
 \textbf{Lookup tables.} As mentioned early, the centroids in SDC are fixed with the setting of $u$, and it can uncompressed represent the recurrent binary vectors. When $u=4$, the distance table in 128-bit registers can be reconstructed directly because the centroids of SDC are presented as 4-bit integers, and the inner product distance range is 8-bit integers. When $u=2$, we use two adjacent dimensions of recurrent binary vector to form 4-bit code, and the distance can be calculated by adding the inner products result of two 2-bit respectively.

 \textbf{SIMD computation.} As the lookup tables and inverted list are prepared, each inverted list is scanned block by block. We depicted this process in Figure \ref{figure_rbe_adc_calculate}. First, the index codes are packed as 8-bit in each cell of 128-bit registers, and we unpacked the subcodes using shifts and masks. For each set of subcodes, the partial distances are yielded using lookup implementation through a combination of shuffle and blends. This process is repeated $u \dot m/4$ times, and the distances are obtained by summing each partial distance with saturated arithmetic. Lastly, each distance is normalized by dividing its magnitude value. In practice, we multiply the distance by the reciprocal of the magnitude value since the multiply operation is fast in SIMD.

\begin{figure}[t]
\centering
\includegraphics[width=0.4\textwidth]{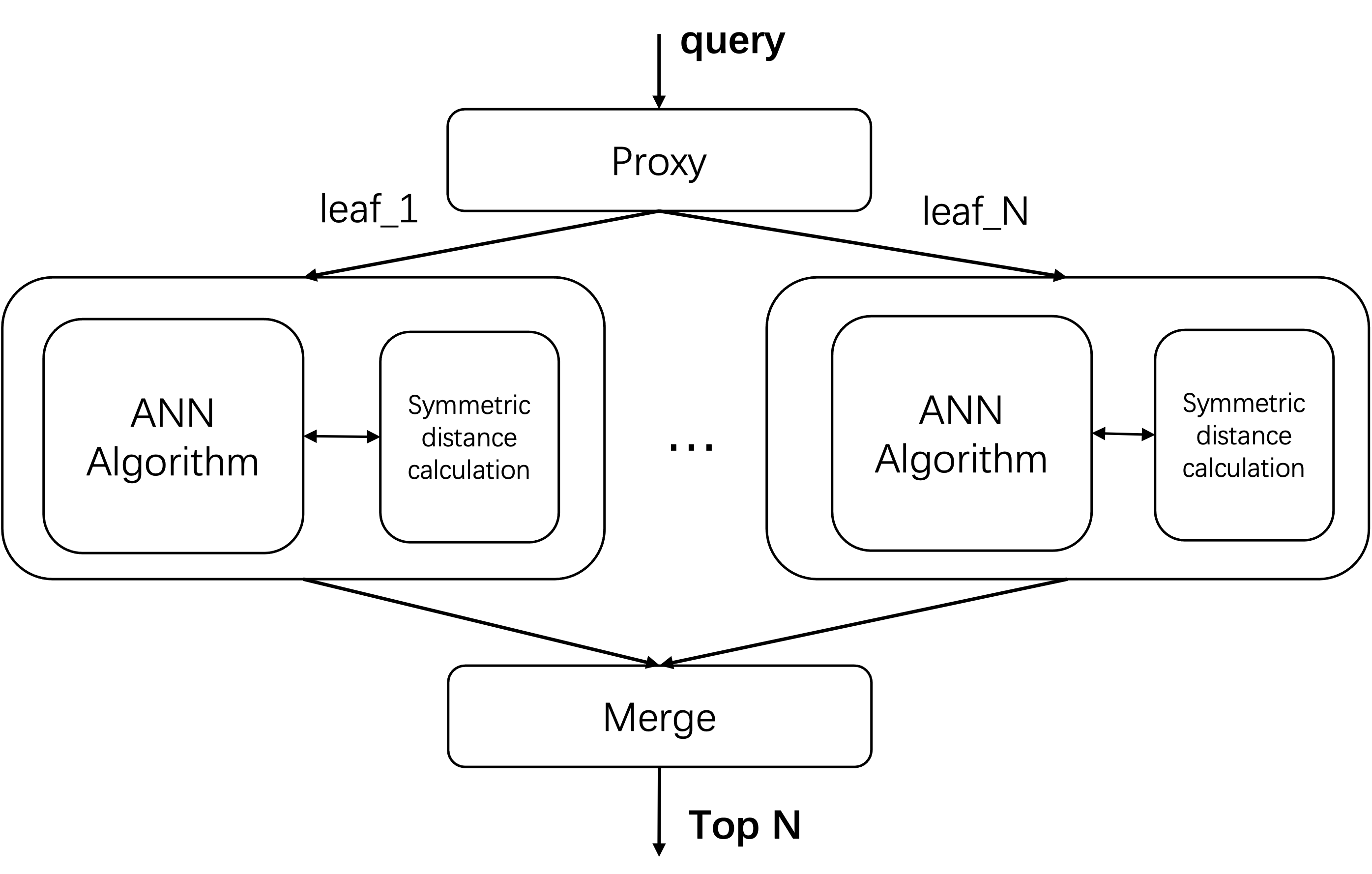} 
\vspace{-5pt}
\caption{Overview of ANN systems equipped with BEBR.}
\vspace{-5pt}
\label{Fig.rbe_system_overview}
\end{figure}

\subsubsection{ANN systems}
We deployed a distributed search system based on ANN search algorithms as in Figure \ref{Fig.rbe_system_overview}. At run time, a query embedding is generated on-the-fly by the embedding learning model. Then, the proxy module dispatches the query to the leaf module, where the main search process happens. Each leaf module equipped various ANN indexes with symmetric distance calculation since our work is orthogonal to ANN algorithms and compatible with any type of index. Therefore, we can choose different algorithms according to the different requirements of the product. For instance, the inverted index (IVF) has two layers for embedding search, one is the coarse layer quantizes embedding vectors into the coarse cluster typically through the $K$-$means$ algorithm, and the other is the fine-grained layer does the efficient calculation of embedding distances. Both layers can be supported by symmetric distance calculation with recurrent binary embeddings used. Lastly, the result from all leaves will be used to produce the top N result through the selection merge process.  


\section{Experiments \label{Sec.exp}}

\subsection{Implementation Details}
The Adam optimizer ~\cite{kingma2014adam} is employed with an initial learning rate of 0.02. The temperature parameter $\tau$ of softmax is set to 0.07. We also adopt gradient clipping when the norm of the gradient exceeds a threshold of 5. 
When training on a server of 8 Nvidia V100 GPUs, the batch size is set to 4096 for binary representation learning and 128 for compatible learning. 
The binarization experiments are based on the PyTorch framework.
We implemented 256-bit SDC in C++, using compiler intrinsics to access SIMD instructions. g++ version 8.1 are selected and enables SSE, AVX, AVX2, and AVX512. Besides, we use Intel MKL 2018 for BLAS. We carry our experiments on Skylake-based servers, which are Tencent cloud instances, built around Intel(R) Xeon(R) Platinum 8255C CPU @ 2.50GHz.

\subsection{Datasets}
We evaluate the proposed BEBR on both public and industrial datasets. For the public dataset, we use image and text data from the MS COCO captioning dataset. For the industrial dataset, we use data collected from two applications. One is web search which returns relevant web pages given a user search query. The other one is video copyright detection which identifies duplicated, replicated, and/or slightly modified versions of a given video sequence (query) in a reference video dataset.
\textbf{Offline datasets:} For web search, we collect search logs of user queries and clicks from \textit{Sogou Search Engine}. After data pre-processing, the training set contains 400 million samples and we use an additional 3 million samples for evaluation. For video copyright detection, we use 8 million images extracted from video sequences to train the model and manually label 30k queries and 600k reference images for validation. Furthermore, we use COCO captioning dataset which contains about 110k 
training images and 5k validation images. For each image in the training and validation set, five independent human-generated captions are provided.
\textbf{Online datasets:} We deploy the proposed BEBR in the production environment of the aforementioned two applications. Web search documents are approximately 6 billion in size, covering the most active web pages on the Internet. The size of image embeddings extracted from video sequences in video copyright detection is about 10 billion.

\subsection{Evaluation Metrics}

\textbf{Offline evaluation.} We use the metric of Recall@k to evaluate the offline performance of the proposed binary-based embedding retrieval method. Specifically, given a query $q$, its relevant documents $\mathcal{D^+} = \{d_1^+, \cdots, d_N^+\}$, and the top-k candidates returned by a model as the retrieval set $\hat{\mathcal{D}} = \{ d_1, \cdots, d_k \}$. $k\gg N$ in practice. Recall@k is defined as:
\begin{equation}
\text{Recall}@k = \frac{|\mathcal{D^+}\cap\hat{\mathcal{D}}|}{N}
\end{equation}

\noindent\textbf{Online Evaluation} We use different metrics to evaluate the effectiveness of our proposed BEBR system.
For web search, we adopt the metrics of click-through rate (CTR) and query rewrite rate (QRR) which are believed to be good indicators of search satisfaction.
For video copyright detection, we conduct the human evaluation for the performance of retrieved documents, Specifically, we ask human evaluators to label the relevance of results from the BEBR system and baseline system. Apart from precision, we report the ratio between the number of copied videos (positive results) and the number of traffic (denoted as detection ratio) to analyze the model's effect on the entire system. A higher detection ratio indicates better performance. 
Furthermore, to evaluate the efficiency of the BEBR system,
We calculate queries per second (QPS) by measuring the amount of search traffic the retrieval stage receives in one second. We also investigate the memory consumption of the search index built in the retrieval stage.

\subsection{Offline Evaluation}

\begin{table}[t]
\centering
\caption{Retrieval performance of different embedding forms on MS COCO caption dataset.}\vspace{-5pt}
\label{Table.binary_public_dataset}
\begin{tabular}{ccccc}
\toprule
  Embedding & Bits & Recall@1 & Recall@5 & Recall@10 \\
\hline
  hash & 1024 & 0.348 & 0.632 & 0.730 \\
  {ours} & {1024} & \textbf{0.360} & \textbf{0.646} & \textbf{0.751} \\
  float & 16384 & \textcolor[RGB]{150,150,150}{0.361} & \textcolor[RGB]{150,150,150}{0.649} & \textcolor[RGB]{150,150,150}{0.744} \\
\bottomrule
\end{tabular}
\end{table}

\begin{table}[t]
\centering
\caption{Retrieval performance of different embedding forms on industrial dataset collected from web search and video copyright detection applications.}\vspace{-5pt}
\label{Table.binary_compare}
\begin{tabular}{ccc}
\toprule
  \multirow{2}{*}{Embedding} & Web search & video copyright detection \\
  \cline{2-3}
  & Recall@10 & Recall@20 \\
\hline
   hash   &  0.819  &   0.688 \\ 
   ours    &  \textbf{0.853}  &  \textbf{0.727} \\
   float  &  \textcolor[RGB]{150,150,150}{0.856}  &   \textcolor[RGB]{150,150,150}{0.734} \\
\bottomrule
\end{tabular}
\end{table}

 \textbf{Effectiveness of recurrent binary embedding.} We investigate the effectiveness of recurrent binary embeddings on both public (academic) and private (industrial) benchmarks. As demonstrated in Tables. \ref{Table.binary_public_dataset}\&\ref{Table.binary_compare}, we compare with the baseline hash \cite{wang2017deep} (1 bit per dimension) and the oracle float (full precision embedding, 32 bits per dimension).

 For the academic dataset, we conduct image-to-text retrieval experiments using the MS COCO caption dataset. Specifically, we employ CLIP ~\cite{radford2021learning} model of ResNet101 to produce float embedding for image and text data. The float embedding of size 16384 bits is then compressed into recurrent binary embedding and hash vector with a size of 1024 bits, achieving a 16x compression ratio. As shown in Table \ref{Table.binary_public_dataset}, recurrent binary embedding surpasses hash embedding and achieves comparable results with float embedding.
 
 For industrial datasets, the vector size of float embeddings in web search and video copyright detection are 8192 and 4096 bits respectively. We adopt the same compression ratio setting of 16x by compressing them into binary embeddings with sizes of 512 and 256 bits respectively. 
 The results are shown in Table \ref{Table.binary_compare}, we achieve comparable retrieval performance with float embedding in web search and video copyright detection applications and surpass hash embedding by 2.4\% and 3.9\% respectively.
 
\begin{table}[t]
\centering
\caption{Comparison with alternative options of binary training pipeline. Experiments are conducted on a web search dataset with 400 million training samples. $\varphi$ denotes recurrent binary model, $\phi$ and $\psi$ denote encoder model for queries and documents.}\vspace{-5pt}
\label{Table.binary_training}
\begin{tabular}{ccc}
\toprule
  Training pipeline & Recall@10 & Training time \\
\hline
   end-to-end  &  0.855 & 125 GPU hours\\
   train $\varphi$ only (fixed $\phi$, $\psi$) & 0.853 & 125 GPU hours \\
   {embedding-to-embedding} & {0.853} & {\textbf{11} GPU hours} \\
\bottomrule
\end{tabular}
\end{table}

\noindent\textbf{Comparison with alternative options of binary training.} 
To investigate the effectiveness and efficiency of our task-agnostic binary training pipeline, we compare it with two alternative options of binary training pipeline. One is end-to-end training where the recurrent binarization module is optimized end-to-end with the backbone network ($\psi$ and $\phi$). The other one adopts a similar pipeline to the end-to-end training but with parameters in $\psi$ and $\phi$ fixed. The fixed $\psi$ and $\phi$ models help improve the retrieval performance by providing data augmentations. Results are shown in Table \ref{Table.binary_training}. Our proposed task-agnostic embedding-to-embedding training pipeline achieves comparable performance with the other two end-to-end training pipelines while reducing training time by 91.2\%.

\begin{table}[t]
\centering
\caption{Comparison with alternative options of backward-compatible training. ($\varphi_{new}$, $\varphi_{old}$) denotes using binary embeddings produced by the new binary model to search binary embeddings produced by the old binary model.}\vspace{-5pt}
\label{Table.bct}
\begin{tabular}{ccc}
\toprule
  Learning strategy  & Comparison pair  & Recall@20 \\
\hline
   baseline  & ($\varphi_{old}$, $\varphi_{old}$) & 0.727 \\
\hline
   normal bct   & ($\varphi_{new}$, $\varphi_{old}$) & 0.765 \\
   two-stage bct & ($\varphi_{new}$, $\varphi_{old}$)   &  0.783 \\
   ours    & ($\varphi_{new}$, $\varphi_{old}$) & \textbf{0.801} \\
\bottomrule
\end{tabular}
\end{table}

\noindent\textbf{Comparison with alternative options of backward compatible training.} 
We investigate the effectiveness of our proposed backward-compatible training pipeline by comparing with two other pipelines. We denote the first alternative pipeline as normal bct where backward compatible training is conducted between new query encoder $\psi_{new}$ and old document encoder $\phi_{old}$. During deployment, backward-compatible binary embeddings are obtained by mapping full precision queries and document embeddings into a common space using the old binary model $\varphi_{old}$. The second alternative pipeline (denote as two-stage bct) contains a two-stage training process where the first stage learns backward compatible full precision embeddings, and the second stage learns backward compatible recurrent binary embeddings based on the compatible output of the first stage.

All experiments of compatible learning are conducted on an offline dataset collected from video copyright detection applications. Results are shown in Table \ref{Table.bct}. All three learning strategies achieve solid backward compatibility by surpassing the baseline where indexing is conducted between the old version of recurrent binary embeddings, indicating the applicability of backward-compatible training in binary embeddings. Among them, our proposed learning paradigm learns better backward compatibility which outperforms normal bct and two-stage bct by 3.6\% and 1.8\%.

\begin{table}[t]
\centering
\caption{Latency for exhaustive search on CPU platform in video copyright detection application.}\vspace{-5pt}
\label{table.latency_cpu_search}
\begin{tabular}{ccccc}
\toprule
  Embedding & Index type & Bits & Search(s)$\downarrow$ & QPS$\uparrow$ \\
\hline
  hash code & bitwise & 256 & 0.0024 & 414  \\
  ours ($u=2$)& bitwise & 256 & 0.0032 & 312  \\
  ours ($u=2$)& SDC & 256 & \textbf{0.0020}  & \textbf{480} \\
  ours ($u=4$)& bitwise &256 & 0.0054 & 185  \\
  ours ($u=4$)& SDC & 256 & \textbf{0.0020}  & \textbf{480}  \\
  float & flat &4096 & 0.05106 & 19  \\
\bottomrule
\end{tabular}
\end{table}

\noindent\textbf{Search latency on CPU platform.} As mentioned in Section \ref{SSec.binary_and_quantization}, the standard search based on recurrent binary embedding \cite{shan2018recurrent} relies on GPU to offer high performance. We implement the standard distance calculation between recurrent binary embeddings on the CPU by using the $popcount$ operation and carry exhaustive search experiment on the offline video copyright detection dataset with recall@20. The experiment loop is, by default, run on a single CPU in a single thread. The comparison between bit-wise operation and SDC results is shown in Table \ref{table.latency_cpu_search}. We observe the bit-wise-based method continues to decrease in QPS with an increase of $u$, and the SDC is almost 2 times faster than the bit-wise operation at $u=4$. Notably, SDC is slightly faster than hash code since the shuffle instructions used in SDC are faster than $popc$.

\begin{figure}[t]
\centering
\includegraphics[width=0.4\textwidth]{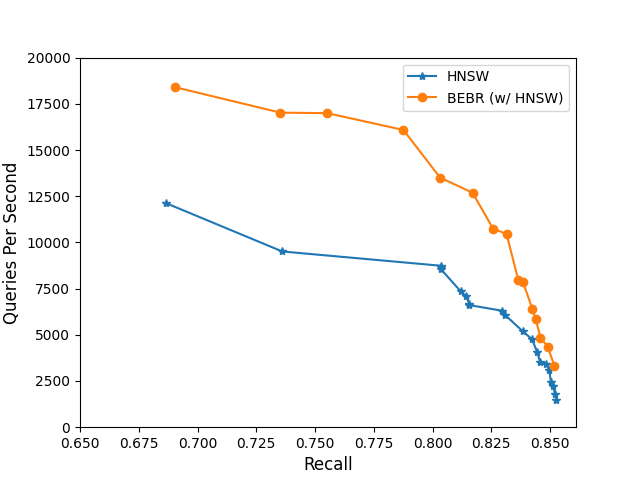} 
\vspace{-5pt}
\caption{ Comparison of retrieval efficiency before and after the deployment of BEBR. Experiments are conducted on an offline dataset collected from web search. }
\vspace{-10pt}
\label{Fig.perf_hnsw}
\end{figure}

\noindent\textbf{Integration of BEBR into ANN algorithms.} Besides the exhaustive search experiments in Table \ref{table.latency_cpu_search}, we also conduct experiments that integrate BEBR into ANN algorithms. Specifically, we equip the HNSW algorithm with the symmetric distance calculation component and leverage recurrent binary embedding for search. Results are illustrated in Figure \ref{Fig.perf_hnsw}. After deploying BEBR, HNSW achieves significant improvements in retrieval efficiency.

\subsection{Online A/B Test}
We deploy the binary embedding-based retrieval system to web search and video copyright detection applications in Tencent and compare them to strong baselines which utilize full precision embedding for retrieval. Note that we substitute full precision embedding with recurrent binary embedding only in the retrieval stage. The subsequent re-rank stages are identical for both settings. Here, we would like to focus on the balance of performance and efficiency, where resource usage is recorded.

The live experiment is conducted over 30\% of the service traffic during one week. Table \ref{Table.ab_test_web} and Table \ref{Table.ab_test_video} show the great benefits of resource and efficiency while retaining performance at the system level. Specifically, BEBR conserves 73.91\% memory usage and increase the QPS of retrieval by 90\%, while CTR and QRR of web search application decrease slightly by 0.02\% and 0.07\% respectively. In video copyright detection, memory usage is reduced by 89.65\%, and QPS is increased by 72\%, while the precision and detection ratio decreases slightly by 0.13\% and 0.21\%. The improvements in retrieval efficiency and storage consumption lead to overall cost reduction. After deploying BEBR, the overall costs of retrieval in web search and video copyright detection are reduced by 55\% and 31\% respectively.


\begin{table}[t]
\centering
\caption{Online A/B tests of BEBR in web search.}\vspace{-5pt}
\label{Table.ab_test_web}
\begin{tabular}{ccccc}
\toprule
  CTR & QRR & Memory usage & QPS\\ 
\hline
  -0.02\%  & -0.07\% & \textbf{-73.91\%} & \textbf{+90\%} \\
\bottomrule
\end{tabular}
\end{table}

\begin{table}[t]
\centering
\caption{Online A/B tests of BEBR in video copyright detection.}\vspace{-5pt}
\label{Table.ab_test_video}
\begin{tabular}{cccc}
\toprule
  Precision & Detection ratio & Memory usage & QPS\\ 
\hline
  -0.13\%  & -0.21\% & \textbf{-89.65\%} & \textbf{+72\%} \\
\bottomrule
\end{tabular}
\end{table}

\section{Conclusion}
The paper presents binary embedding-based retrieval (BEBR) to improve retrieval efficiency and reduce storage consumption while retaining retrieval performance in Tencent products.
Specifically, we 
1) compress full-precision embedding into recurrent binary embedding using a lightweight transformation model; 2) adopt a new task-agnostic embedding-to-embedding strategy to enable efficient training and deployment of binary embeddings; 3) investigate backward-compatible training in binary embeddings to enable refresh-free embedding model upgrades; 4) propose symmetric distance calculation equipped with ANN algorithms to form an efficient index system. BEBR has been successfully deployed into Tencent's products, including web search (Sogou), QQ, and Tencent Video. We hope our work can well inspire the community to effectively deliver research achievements into real-world applications.

\begin{acks}
We sincerely appreciate all the colleagues in the project of BEBR development, including but not limited to Yang Li, Qiugen Xiao, Dezhang Yuan, Yang Fang, Chen Xu, and Xiaohu Qie, for their valuable discussions, efforts, and support.
We thank all the reviewers and chairs for their time and constructive comments.
\end{acks}

\bibliographystyle{ACM-Reference-Format}
\bibliography{template/acmart}


\begin{thebibliography}{54}


\ifx \showCODEN    \undefined \def \showCODEN     #1{\unskip}     \fi
\ifx \showDOI      \undefined \def \showDOI       #1{#1}\fi
\ifx \showISBNx    \undefined \def \showISBNx     #1{\unskip}     \fi
\ifx \showISBNxiii \undefined \def \showISBNxiii  #1{\unskip}     \fi
\ifx \showISSN     \undefined \def \showISSN      #1{\unskip}     \fi
\ifx \showLCCN     \undefined \def \showLCCN      #1{\unskip}     \fi
\ifx \shownote     \undefined \def \shownote      #1{#1}          \fi
\ifx \showarticletitle \undefined \def \showarticletitle #1{#1}   \fi
\ifx \showURL      \undefined \def \showURL       {\relax}        \fi
\providecommand\bibfield[2]{#2}
\providecommand\bibinfo[2]{#2}
\providecommand\natexlab[1]{#1}
\providecommand\showeprint[2][]{arXiv:#2}

\bibitem[Agarap(2018)]%
        {agarap2018deep}
\bibfield{author}{\bibinfo{person}{Abien~Fred Agarap}.}
  \bibinfo{year}{2018}\natexlab{}.
\newblock \showarticletitle{Deep learning using rectified linear units (relu)}.
\newblock \bibinfo{journal}{\emph{arXiv preprint arXiv:1803.08375}}
  (\bibinfo{year}{2018}).
\newblock


\bibitem[Andr{\'e} et~al\mbox{.}(2017)]%
        {andre2017accelerated}
\bibfield{author}{\bibinfo{person}{Fabien Andr{\'e}},
  \bibinfo{person}{Anne-Marie Kermarrec}, {and} \bibinfo{person}{Nicolas
  Le~Scouarnec}.} \bibinfo{year}{2017}\natexlab{}.
\newblock \showarticletitle{Accelerated nearest neighbor search with quick
  adc}. In \bibinfo{booktitle}{\emph{Proceedings of the 2017 ACM on
  International Conference on Multimedia Retrieval}}.
  \bibinfo{pages}{159--166}.
\newblock


\bibitem[Andr{\'e} et~al\mbox{.}(2019a)]%
        {andre2019quicker}
\bibfield{author}{\bibinfo{person}{Fabien Andr{\'e}},
  \bibinfo{person}{Anne-Marie Kermarrec}, {and} \bibinfo{person}{Nicolas
  Le~Scouarnec}.} \bibinfo{year}{2019}\natexlab{a}.
\newblock \showarticletitle{Quicker adc: Unlocking the hidden potential of
  product quantization with simd}.
\newblock \bibinfo{journal}{\emph{IEEE transactions on pattern analysis and
  machine intelligence}} (\bibinfo{year}{2019}), \bibinfo{pages}{1666--1677}.
\newblock


\bibitem[Andr{\'e} et~al\mbox{.}(2019b)]%
        {andre2019derived}
\bibfield{author}{\bibinfo{person}{Fabien Andr{\'e}},
  \bibinfo{person}{Anne-Marie Kermarrec}, {and} \bibinfo{person}{Nicolas~Le
  Scouarnec}.} \bibinfo{year}{2019}\natexlab{b}.
\newblock \showarticletitle{Derived codebooks for high-accuracy nearest
  neighbor search}.
\newblock \bibinfo{journal}{\emph{arXiv preprint arXiv:1905.06900}}
  (\bibinfo{year}{2019}).
\newblock


\bibitem[Babenko and Lempitsky(2014)]%
        {babenko2014additive}
\bibfield{author}{\bibinfo{person}{Artem Babenko} {and} \bibinfo{person}{Victor
  Lempitsky}.} \bibinfo{year}{2014}\natexlab{}.
\newblock \showarticletitle{Additive quantization for extreme vector
  compression}. In \bibinfo{booktitle}{\emph{Proceedings of the IEEE Conference
  on Computer Vision and Pattern Recognition}}. \bibinfo{pages}{931--938}.
\newblock


\bibitem[Babenko and Lempitsky(2015)]%
        {babenko2015tree}
\bibfield{author}{\bibinfo{person}{Artem Babenko} {and} \bibinfo{person}{Victor
  Lempitsky}.} \bibinfo{year}{2015}\natexlab{}.
\newblock \showarticletitle{Tree quantization for large-scale similarity search
  and classification}. In \bibinfo{booktitle}{\emph{Proceedings of the IEEE
  Conference on Computer Vision and Pattern Recognition}}.
  \bibinfo{pages}{4240--4248}.
\newblock


\bibitem[Blalock and Guttag(2017)]%
        {blalock2017bolt}
\bibfield{author}{\bibinfo{person}{Davis~W Blalock} {and}
  \bibinfo{person}{John~V Guttag}.} \bibinfo{year}{2017}\natexlab{}.
\newblock \showarticletitle{Bolt: Accelerated data mining with fast vector
  compression}. In \bibinfo{booktitle}{\emph{Proceedings of the 23rd ACM SIGKDD
  International Conference on Knowledge Discovery and Data Mining}}.
  \bibinfo{pages}{727--735}.
\newblock


\bibitem[Broder et~al\mbox{.}(1997)]%
        {broder1997syntactic}
\bibfield{author}{\bibinfo{person}{Andrei~Z Broder}, \bibinfo{person}{Steven~C
  Glassman}, \bibinfo{person}{Mark~S Manasse}, {and} \bibinfo{person}{Geoffrey
  Zweig}.} \bibinfo{year}{1997}\natexlab{}.
\newblock \showarticletitle{Syntactic clustering of the web}.
\newblock \bibinfo{journal}{\emph{Computer networks and ISDN systems}}
  \bibinfo{volume}{29}, \bibinfo{number}{8-13} (\bibinfo{year}{1997}),
  \bibinfo{pages}{1157--1166}.
\newblock


\bibitem[Budnik and Avrithis(2021)]%
        {budnik2021asymmetric}
\bibfield{author}{\bibinfo{person}{Mateusz Budnik} {and}
  \bibinfo{person}{Yannis Avrithis}.} \bibinfo{year}{2021}\natexlab{}.
\newblock \showarticletitle{Asymmetric metric learning for knowledge transfer}.
  In \bibinfo{booktitle}{\emph{Proceedings of the IEEE/CVF Conference on
  Computer Vision and Pattern Recognition}}. \bibinfo{pages}{8228--8238}.
\newblock


\bibitem[Cao et~al\mbox{.}(2017)]%
        {cao2017hashnet}
\bibfield{author}{\bibinfo{person}{Zhangjie Cao}, \bibinfo{person}{Mingsheng
  Long}, \bibinfo{person}{Jianmin Wang}, {and} \bibinfo{person}{Philip~S Yu}.}
  \bibinfo{year}{2017}\natexlab{}.
\newblock \showarticletitle{Hashnet: Deep learning to hash by continuation}. In
  \bibinfo{booktitle}{\emph{Proceedings of the IEEE international conference on
  computer vision}}. \bibinfo{pages}{5608--5617}.
\newblock


\bibitem[Charikar(2002)]%
        {charikar2002similarity}
\bibfield{author}{\bibinfo{person}{Moses~S Charikar}.}
  \bibinfo{year}{2002}\natexlab{}.
\newblock \showarticletitle{Similarity estimation techniques from rounding
  algorithms}. In \bibinfo{booktitle}{\emph{Proceedings of the thiry-fourth
  annual ACM symposium on Theory of computing}}. \bibinfo{pages}{380--388}.
\newblock


\bibitem[Chen et~al\mbox{.}(2019)]%
        {chen2019r3}
\bibfield{author}{\bibinfo{person}{Ken Chen}, \bibinfo{person}{Yichao Wu},
  \bibinfo{person}{Haoyu Qin}, \bibinfo{person}{Ding Liang},
  \bibinfo{person}{Xuebo Liu}, {and} \bibinfo{person}{Junjie Yan}.}
  \bibinfo{year}{2019}\natexlab{}.
\newblock \showarticletitle{R3 adversarial network for cross model face
  recognition}. In \bibinfo{booktitle}{\emph{Proceedings of the IEEE/CVF
  Conference on Computer Vision and Pattern Recognition}}.
  \bibinfo{pages}{9868--9876}.
\newblock


\bibitem[Chen et~al\mbox{.}(2020)]%
        {chen2020simple}
\bibfield{author}{\bibinfo{person}{Ting Chen}, \bibinfo{person}{Simon
  Kornblith}, \bibinfo{person}{Mohammad Norouzi}, {and}
  \bibinfo{person}{Geoffrey Hinton}.} \bibinfo{year}{2020}\natexlab{}.
\newblock \showarticletitle{A simple framework for contrastive learning of
  visual representations}. In \bibinfo{booktitle}{\emph{International
  conference on machine learning}}. PMLR, \bibinfo{pages}{1597--1607}.
\newblock


\bibitem[Courbariaux et~al\mbox{.}(2016)]%
        {courbariaux2016binarized}
\bibfield{author}{\bibinfo{person}{Matthieu Courbariaux}, \bibinfo{person}{Itay
  Hubara}, \bibinfo{person}{Daniel Soudry}, \bibinfo{person}{Ran El-Yaniv},
  {and} \bibinfo{person}{Yoshua Bengio}.} \bibinfo{year}{2016}\natexlab{}.
\newblock \showarticletitle{Binarized neural networks: Training deep neural
  networks with weights and activations constrained to+ 1 or-1}.
\newblock \bibinfo{journal}{\emph{arXiv preprint arXiv:1602.02830}}
  (\bibinfo{year}{2016}).
\newblock


\bibitem[Devlin et~al\mbox{.}(2018)]%
        {devlin2018bert}
\bibfield{author}{\bibinfo{person}{Jacob Devlin}, \bibinfo{person}{Ming-Wei
  Chang}, \bibinfo{person}{Kenton Lee}, {and} \bibinfo{person}{Kristina
  Toutanova}.} \bibinfo{year}{2018}\natexlab{}.
\newblock \showarticletitle{Bert: Pre-training of deep bidirectional
  transformers for language understanding}.
\newblock \bibinfo{journal}{\emph{arXiv preprint arXiv:1810.04805}}
  (\bibinfo{year}{2018}).
\newblock


\bibitem[Duggal et~al\mbox{.}(2021)]%
        {duggal2021compatibility}
\bibfield{author}{\bibinfo{person}{Rahul Duggal}, \bibinfo{person}{Hao Zhou},
  \bibinfo{person}{Shuo Yang}, \bibinfo{person}{Yuanjun Xiong},
  \bibinfo{person}{Wei Xia}, \bibinfo{person}{Zhuowen Tu}, {and}
  \bibinfo{person}{Stefano Soatto}.} \bibinfo{year}{2021}\natexlab{}.
\newblock \showarticletitle{Compatibility-aware heterogeneous visual search}.
  In \bibinfo{booktitle}{\emph{Proceedings of the IEEE/CVF Conference on
  Computer Vision and Pattern Recognition}}. \bibinfo{pages}{10723--10732}.
\newblock


\bibitem[Fan et~al\mbox{.}(2020)]%
        {fan2020deep}
\bibfield{author}{\bibinfo{person}{Lixin Fan}, \bibinfo{person}{Kam~Woh Ng},
  \bibinfo{person}{Ce Ju}, \bibinfo{person}{Tianyu Zhang}, {and}
  \bibinfo{person}{Chee~Seng Chan}.} \bibinfo{year}{2020}\natexlab{}.
\newblock \showarticletitle{Deep Polarized Network for Supervised Learning of
  Accurate Binary Hashing Codes.}. In \bibinfo{booktitle}{\emph{IJCAI}}.
  \bibinfo{pages}{825--831}.
\newblock


\bibitem[Fu et~al\mbox{.}(2017)]%
        {fu2017fast}
\bibfield{author}{\bibinfo{person}{Cong Fu}, \bibinfo{person}{Chao Xiang},
  \bibinfo{person}{Changxu Wang}, {and} \bibinfo{person}{Deng Cai}.}
  \bibinfo{year}{2017}\natexlab{}.
\newblock \showarticletitle{Fast approximate nearest neighbor search with the
  navigating spreading-out graph}.
\newblock \bibinfo{journal}{\emph{arXiv preprint arXiv:1707.00143}}
  (\bibinfo{year}{2017}).
\newblock


\bibitem[Gan et~al\mbox{.}(2012)]%
        {gan2012locality}
\bibfield{author}{\bibinfo{person}{Junhao Gan}, \bibinfo{person}{Jianlin Feng},
  \bibinfo{person}{Qiong Fang}, {and} \bibinfo{person}{Wilfred Ng}.}
  \bibinfo{year}{2012}\natexlab{}.
\newblock \showarticletitle{Locality-sensitive hashing scheme based on dynamic
  collision counting}. In \bibinfo{booktitle}{\emph{Proceedings of the 2012 ACM
  SIGMOD international conference on management of data}}.
  \bibinfo{pages}{541--552}.
\newblock


\bibitem[Ge et~al\mbox{.}(2013)]%
        {ge2013optimized}
\bibfield{author}{\bibinfo{person}{Tiezheng Ge}, \bibinfo{person}{Kaiming He},
  \bibinfo{person}{Qifa Ke}, {and} \bibinfo{person}{Jian Sun}.}
  \bibinfo{year}{2013}\natexlab{}.
\newblock \showarticletitle{Optimized product quantization}.
\newblock \bibinfo{journal}{\emph{IEEE transactions on pattern analysis and
  machine intelligence}} \bibinfo{volume}{36}, \bibinfo{number}{4}
  (\bibinfo{year}{2013}), \bibinfo{pages}{744--755}.
\newblock


\bibitem[Guo et~al\mbox{.}(2022)]%
        {guo2022semantic}
\bibfield{author}{\bibinfo{person}{Jiafeng Guo}, \bibinfo{person}{Yinqiong
  Cai}, \bibinfo{person}{Yixing Fan}, \bibinfo{person}{Fei Sun},
  \bibinfo{person}{Ruqing Zhang}, {and} \bibinfo{person}{Xueqi Cheng}.}
  \bibinfo{year}{2022}\natexlab{}.
\newblock \showarticletitle{Semantic models for the first-stage retrieval: A
  comprehensive review}.
\newblock \bibinfo{journal}{\emph{ACM Transactions on Information Systems
  (TOIS)}} \bibinfo{volume}{40}, \bibinfo{number}{4} (\bibinfo{year}{2022}),
  \bibinfo{pages}{1--42}.
\newblock


\bibitem[He et~al\mbox{.}(2020)]%
        {he2020momentum}
\bibfield{author}{\bibinfo{person}{Kaiming He}, \bibinfo{person}{Haoqi Fan},
  \bibinfo{person}{Yuxin Wu}, \bibinfo{person}{Saining Xie}, {and}
  \bibinfo{person}{Ross Girshick}.} \bibinfo{year}{2020}\natexlab{}.
\newblock \showarticletitle{Momentum contrast for unsupervised visual
  representation learning}. In \bibinfo{booktitle}{\emph{Proceedings of the
  IEEE/CVF conference on computer vision and pattern recognition}}.
  \bibinfo{pages}{9729--9738}.
\newblock


\bibitem[He et~al\mbox{.}(2016)]%
        {he2016deep}
\bibfield{author}{\bibinfo{person}{Kaiming He}, \bibinfo{person}{Xiangyu
  Zhang}, \bibinfo{person}{Shaoqing Ren}, {and} \bibinfo{person}{Jian Sun}.}
  \bibinfo{year}{2016}\natexlab{}.
\newblock \showarticletitle{Deep residual learning for image recognition}. In
  \bibinfo{booktitle}{\emph{Proceedings of the IEEE conference on computer
  vision and pattern recognition}}. \bibinfo{pages}{770--778}.
\newblock


\bibitem[Hu et~al\mbox{.}(2022)]%
        {hu2022learning}
\bibfield{author}{\bibinfo{person}{Weihua Hu}, \bibinfo{person}{Rajas Bansal},
  \bibinfo{person}{Kaidi Cao}, \bibinfo{person}{Nikhil Rao},
  \bibinfo{person}{Karthik Subbian}, {and} \bibinfo{person}{Jure Leskovec}.}
  \bibinfo{year}{2022}\natexlab{}.
\newblock \showarticletitle{Learning Backward Compatible Embeddings}.
\newblock \bibinfo{journal}{\emph{arXiv preprint arXiv:2206.03040}}
  (\bibinfo{year}{2022}).
\newblock


\bibitem[Huang et~al\mbox{.}(2020)]%
        {huang2020embedding}
\bibfield{author}{\bibinfo{person}{Jui-Ting Huang}, \bibinfo{person}{Ashish
  Sharma}, \bibinfo{person}{Shuying Sun}, \bibinfo{person}{Li Xia},
  \bibinfo{person}{David Zhang}, \bibinfo{person}{Philip Pronin},
  \bibinfo{person}{Janani Padmanabhan}, \bibinfo{person}{Giuseppe Ottaviano},
  {and} \bibinfo{person}{Linjun Yang}.} \bibinfo{year}{2020}\natexlab{}.
\newblock \showarticletitle{Embedding-based retrieval in facebook search}. In
  \bibinfo{booktitle}{\emph{Proceedings of the 26th ACM SIGKDD International
  Conference on Knowledge Discovery \& Data Mining}}.
  \bibinfo{pages}{2553--2561}.
\newblock


\bibitem[Indyk and Motwani(1998)]%
        {indyk1998approximate}
\bibfield{author}{\bibinfo{person}{Piotr Indyk} {and} \bibinfo{person}{Rajeev
  Motwani}.} \bibinfo{year}{1998}\natexlab{}.
\newblock \showarticletitle{Approximate nearest neighbors: towards removing the
  curse of dimensionality}. In \bibinfo{booktitle}{\emph{Proceedings of the
  thirtieth annual ACM symposium on Theory of computing}}.
  \bibinfo{pages}{604--613}.
\newblock


\bibitem[Jegou et~al\mbox{.}(2010)]%
        {jegou2010product}
\bibfield{author}{\bibinfo{person}{Herve Jegou}, \bibinfo{person}{Matthijs
  Douze}, {and} \bibinfo{person}{Cordelia Schmid}.}
  \bibinfo{year}{2010}\natexlab{}.
\newblock \showarticletitle{Product quantization for nearest neighbor search}.
\newblock \bibinfo{journal}{\emph{IEEE transactions on pattern analysis and
  machine intelligence}} \bibinfo{volume}{33}, \bibinfo{number}{1}
  (\bibinfo{year}{2010}), \bibinfo{pages}{117--128}.
\newblock


\bibitem[Ji et~al\mbox{.}(2012)]%
        {ji2012super}
\bibfield{author}{\bibinfo{person}{Jianqiu Ji}, \bibinfo{person}{Jianmin Li},
  \bibinfo{person}{Shuicheng Yan}, \bibinfo{person}{Bo Zhang}, {and}
  \bibinfo{person}{Qi Tian}.} \bibinfo{year}{2012}\natexlab{}.
\newblock \showarticletitle{Super-bit locality-sensitive hashing}.
\newblock \bibinfo{journal}{\emph{Advances in neural information processing
  systems}}  \bibinfo{volume}{25} (\bibinfo{year}{2012}).
\newblock


\bibitem[Kingma and Ba(2014)]%
        {kingma2014adam}
\bibfield{author}{\bibinfo{person}{Diederik~P Kingma} {and}
  \bibinfo{person}{Jimmy Ba}.} \bibinfo{year}{2014}\natexlab{}.
\newblock \showarticletitle{Adam: A method for stochastic optimization}.
\newblock \bibinfo{journal}{\emph{arXiv preprint arXiv:1412.6980}}
  (\bibinfo{year}{2014}).
\newblock


\bibitem[Li et~al\mbox{.}(2006)]%
        {li2006very}
\bibfield{author}{\bibinfo{person}{Ping Li}, \bibinfo{person}{Trevor~J Hastie},
  {and} \bibinfo{person}{Kenneth~W Church}.} \bibinfo{year}{2006}\natexlab{}.
\newblock \showarticletitle{Very sparse random projections}. In
  \bibinfo{booktitle}{\emph{Proceedings of the 12th ACM SIGKDD international
  conference on Knowledge discovery and data mining}}.
  \bibinfo{pages}{287--296}.
\newblock


\bibitem[Li et~al\mbox{.}(2021)]%
        {li2021embedding}
\bibfield{author}{\bibinfo{person}{Sen Li}, \bibinfo{person}{Fuyu Lv},
  \bibinfo{person}{Taiwei Jin}, \bibinfo{person}{Guli Lin},
  \bibinfo{person}{Keping Yang}, \bibinfo{person}{Xiaoyi Zeng},
  \bibinfo{person}{Xiao-Ming Wu}, {and} \bibinfo{person}{Qianli Ma}.}
  \bibinfo{year}{2021}\natexlab{}.
\newblock \showarticletitle{Embedding-based product retrieval in taobao
  search}. In \bibinfo{booktitle}{\emph{Proceedings of the 27th ACM SIGKDD
  Conference on Knowledge Discovery \& Data Mining}}.
  \bibinfo{pages}{3181--3189}.
\newblock


\bibitem[Li et~al\mbox{.}(2015)]%
        {li2015feature}
\bibfield{author}{\bibinfo{person}{Wu-Jun Li}, \bibinfo{person}{Sheng Wang},
  {and} \bibinfo{person}{Wang-Cheng Kang}.} \bibinfo{year}{2015}\natexlab{}.
\newblock \showarticletitle{Feature learning based deep supervised hashing with
  pairwise labels}.
\newblock \bibinfo{journal}{\emph{arXiv preprint arXiv:1511.03855}}
  (\bibinfo{year}{2015}).
\newblock


\bibitem[Liu et~al\mbox{.}(2018)]%
        {liu2018deep}
\bibfield{author}{\bibinfo{person}{Bin Liu}, \bibinfo{person}{Yue Cao},
  \bibinfo{person}{Mingsheng Long}, \bibinfo{person}{Jianmin Wang}, {and}
  \bibinfo{person}{Jingdong Wang}.} \bibinfo{year}{2018}\natexlab{}.
\newblock \showarticletitle{Deep triplet quantization}. In
  \bibinfo{booktitle}{\emph{Proceedings of the 26th ACM international
  conference on Multimedia}}. \bibinfo{pages}{755--763}.
\newblock


\bibitem[Malkov et~al\mbox{.}(2012)]%
        {malkov2012scalable}
\bibfield{author}{\bibinfo{person}{Yury Malkov}, \bibinfo{person}{Alexander
  Ponomarenko}, \bibinfo{person}{Andrey Logvinov}, {and}
  \bibinfo{person}{Vladimir Krylov}.} \bibinfo{year}{2012}\natexlab{}.
\newblock \showarticletitle{Scalable distributed algorithm for approximate
  nearest neighbor search problem in high dimensional general metric spaces}.
  In \bibinfo{booktitle}{\emph{Similarity Search and Applications: 5th
  International Conference, SISAP 2012, Toronto, ON, Canada, August 9-10, 2012.
  Proceedings 5}}. Springer, \bibinfo{pages}{132--147}.
\newblock


\bibitem[Malkov and Yashunin(2018)]%
        {malkov2018efficient}
\bibfield{author}{\bibinfo{person}{Yu~A Malkov} {and} \bibinfo{person}{Dmitry~A
  Yashunin}.} \bibinfo{year}{2018}\natexlab{}.
\newblock \showarticletitle{Efficient and robust approximate nearest neighbor
  search using hierarchical navigable small world graphs}.
\newblock \bibinfo{journal}{\emph{IEEE transactions on pattern analysis and
  machine intelligence}} \bibinfo{volume}{42}, \bibinfo{number}{4}
  (\bibinfo{year}{2018}), \bibinfo{pages}{824--836}.
\newblock


\bibitem[Meng et~al\mbox{.}(2021)]%
        {meng2021learning}
\bibfield{author}{\bibinfo{person}{Qiang Meng}, \bibinfo{person}{Chixiang
  Zhang}, \bibinfo{person}{Xiaoqiang Xu}, {and} \bibinfo{person}{Feng Zhou}.}
  \bibinfo{year}{2021}\natexlab{}.
\newblock \showarticletitle{Learning compatible embeddings}. In
  \bibinfo{booktitle}{\emph{Proceedings of the IEEE/CVF International
  Conference on Computer Vision}}. \bibinfo{pages}{9939--9948}.
\newblock


\bibitem[Moffat and Zobel(1996)]%
        {moffat1996self}
\bibfield{author}{\bibinfo{person}{Alistair Moffat} {and}
  \bibinfo{person}{Justin Zobel}.} \bibinfo{year}{1996}\natexlab{}.
\newblock \showarticletitle{Self-indexing inverted files for fast text
  retrieval}.
\newblock \bibinfo{journal}{\emph{ACM Transactions on Information Systems
  (TOIS)}} \bibinfo{volume}{14}, \bibinfo{number}{4} (\bibinfo{year}{1996}),
  \bibinfo{pages}{349--379}.
\newblock


\bibitem[Nigam et~al\mbox{.}(2019)]%
        {nigam2019semantic}
\bibfield{author}{\bibinfo{person}{Priyanka Nigam}, \bibinfo{person}{Yiwei
  Song}, \bibinfo{person}{Vijai Mohan}, \bibinfo{person}{Vihan Lakshman},
  \bibinfo{person}{Weitian Ding}, \bibinfo{person}{Ankit Shingavi},
  \bibinfo{person}{Choon~Hui Teo}, \bibinfo{person}{Hao Gu}, {and}
  \bibinfo{person}{Bing Yin}.} \bibinfo{year}{2019}\natexlab{}.
\newblock \showarticletitle{Semantic product search}. In
  \bibinfo{booktitle}{\emph{Proceedings of the 25th ACM SIGKDD International
  Conference on Knowledge Discovery \& Data Mining}}.
  \bibinfo{pages}{2876--2885}.
\newblock


\bibitem[Norouzi and Fleet(2013)]%
        {norouzi2013cartesian}
\bibfield{author}{\bibinfo{person}{Mohammad Norouzi} {and}
  \bibinfo{person}{David~J Fleet}.} \bibinfo{year}{2013}\natexlab{}.
\newblock \showarticletitle{Cartesian k-means}. In
  \bibinfo{booktitle}{\emph{Proceedings of the IEEE Conference on computer
  Vision and Pattern Recognition}}. \bibinfo{pages}{3017--3024}.
\newblock


\bibitem[Ou et~al\mbox{.}(2021)]%
        {ou2021refining}
\bibfield{author}{\bibinfo{person}{Zijing Ou}, \bibinfo{person}{Qinliang Su},
  \bibinfo{person}{Jianxing Yu}, \bibinfo{person}{Ruihui Zhao},
  \bibinfo{person}{Yefeng Zheng}, {and} \bibinfo{person}{Bang Liu}.}
  \bibinfo{year}{2021}\natexlab{}.
\newblock \showarticletitle{Refining BERT Embeddings for Document Hashing via
  Mutual Information Maximization}.
\newblock \bibinfo{journal}{\emph{arXiv preprint arXiv:2109.02867}}
  (\bibinfo{year}{2021}).
\newblock


\bibitem[Radford et~al\mbox{.}(2021)]%
        {radford2021learning}
\bibfield{author}{\bibinfo{person}{Alec Radford}, \bibinfo{person}{Jong~Wook
  Kim}, \bibinfo{person}{Chris Hallacy}, \bibinfo{person}{Aditya Ramesh},
  \bibinfo{person}{Gabriel Goh}, \bibinfo{person}{Sandhini Agarwal},
  \bibinfo{person}{Girish Sastry}, \bibinfo{person}{Amanda Askell},
  \bibinfo{person}{Pamela Mishkin}, \bibinfo{person}{Jack Clark},
  {et~al\mbox{.}}} \bibinfo{year}{2021}\natexlab{}.
\newblock \showarticletitle{Learning transferable visual models from natural
  language supervision}. In \bibinfo{booktitle}{\emph{International conference
  on machine learning}}. PMLR, \bibinfo{pages}{8748--8763}.
\newblock


\bibitem[Robertson et~al\mbox{.}(2009)]%
        {robertson2009probabilistic}
\bibfield{author}{\bibinfo{person}{Stephen Robertson}, \bibinfo{person}{Hugo
  Zaragoza}, {et~al\mbox{.}}} \bibinfo{year}{2009}\natexlab{}.
\newblock \showarticletitle{The probabilistic relevance framework: BM25 and
  beyond}.
\newblock \bibinfo{journal}{\emph{Foundations and Trends{\textregistered} in
  Information Retrieval}} \bibinfo{volume}{3}, \bibinfo{number}{4}
  (\bibinfo{year}{2009}), \bibinfo{pages}{333--389}.
\newblock


\bibitem[Schroff et~al\mbox{.}(2015)]%
        {schroff2015facenet}
\bibfield{author}{\bibinfo{person}{Florian Schroff}, \bibinfo{person}{Dmitry
  Kalenichenko}, {and} \bibinfo{person}{James Philbin}.}
  \bibinfo{year}{2015}\natexlab{}.
\newblock \showarticletitle{Facenet: A unified embedding for face recognition
  and clustering}. In \bibinfo{booktitle}{\emph{Proceedings of the IEEE
  conference on computer vision and pattern recognition}}.
  \bibinfo{pages}{815--823}.
\newblock


\bibitem[Shan et~al\mbox{.}(2018)]%
        {shan2018recurrent}
\bibfield{author}{\bibinfo{person}{Ying Shan}, \bibinfo{person}{Jian Jiao},
  \bibinfo{person}{Jie Zhu}, {and} \bibinfo{person}{JC Mao}.}
  \bibinfo{year}{2018}\natexlab{}.
\newblock \showarticletitle{Recurrent binary embedding for gpu-enabled
  exhaustive retrieval from billion-scale semantic vectors}. In
  \bibinfo{booktitle}{\emph{Proceedings of the 24th ACM SIGKDD International
  Conference on Knowledge Discovery \& Data Mining}}.
  \bibinfo{pages}{2170--2179}.
\newblock


\bibitem[Shen et~al\mbox{.}(2020)]%
        {shen2020towards}
\bibfield{author}{\bibinfo{person}{Yantao Shen}, \bibinfo{person}{Yuanjun
  Xiong}, \bibinfo{person}{Wei Xia}, {and} \bibinfo{person}{Stefano Soatto}.}
  \bibinfo{year}{2020}\natexlab{}.
\newblock \showarticletitle{Towards backward-compatible representation
  learning}. In \bibinfo{booktitle}{\emph{Proceedings of the IEEE/CVF
  Conference on Computer Vision and Pattern Recognition}}.
  \bibinfo{pages}{6368--6377}.
\newblock


\bibitem[Shrivastava et~al\mbox{.}(2016)]%
        {shrivastava2016training}
\bibfield{author}{\bibinfo{person}{Abhinav Shrivastava},
  \bibinfo{person}{Abhinav Gupta}, {and} \bibinfo{person}{Ross Girshick}.}
  \bibinfo{year}{2016}\natexlab{}.
\newblock \showarticletitle{Training region-based object detectors with online
  hard example mining}. In \bibinfo{booktitle}{\emph{Proceedings of the IEEE
  conference on computer vision and pattern recognition}}.
  \bibinfo{pages}{761--769}.
\newblock


\bibitem[Su et~al\mbox{.}(2018)]%
        {su2018greedy}
\bibfield{author}{\bibinfo{person}{Shupeng Su}, \bibinfo{person}{Chao Zhang},
  \bibinfo{person}{Kai Han}, {and} \bibinfo{person}{Yonghong Tian}.}
  \bibinfo{year}{2018}\natexlab{}.
\newblock \showarticletitle{Greedy hash: Towards fast optimization for accurate
  hash coding in cnn}.
\newblock \bibinfo{journal}{\emph{Advances in neural information processing
  systems}}  \bibinfo{volume}{31} (\bibinfo{year}{2018}).
\newblock


\bibitem[Wang et~al\mbox{.}(2020)]%
        {wang2020unified}
\bibfield{author}{\bibinfo{person}{Chien-Yi Wang}, \bibinfo{person}{Ya-Liang
  Chang}, \bibinfo{person}{Shang-Ta Yang}, \bibinfo{person}{Dong Chen}, {and}
  \bibinfo{person}{Shang-Hong Lai}.} \bibinfo{year}{2020}\natexlab{}.
\newblock \showarticletitle{Unified representation learning for cross model
  compatibility}.
\newblock \bibinfo{journal}{\emph{arXiv preprint arXiv:2008.04821}}
  (\bibinfo{year}{2020}).
\newblock


\bibitem[Wang et~al\mbox{.}(2017)]%
        {wang2017deep}
\bibfield{author}{\bibinfo{person}{Xiaofang Wang}, \bibinfo{person}{Yi Shi},
  {and} \bibinfo{person}{Kris~M Kitani}.} \bibinfo{year}{2017}\natexlab{}.
\newblock \showarticletitle{Deep supervised hashing with triplet labels}. In
  \bibinfo{booktitle}{\emph{Computer Vision--ACCV 2016: 13th Asian Conference
  on Computer Vision, Taipei, Taiwan, November 20-24, 2016, Revised Selected
  Papers, Part I 13}}. Springer, \bibinfo{pages}{70--84}.
\newblock


\bibitem[Wu et~al\mbox{.}(2020)]%
        {wu2020zero}
\bibfield{author}{\bibinfo{person}{Tao Wu}, \bibinfo{person}{Ellie Ka-In Chio},
  \bibinfo{person}{Heng-Tze Cheng}, \bibinfo{person}{Yu Du},
  \bibinfo{person}{Steffen Rendle}, \bibinfo{person}{Dima Kuzmin},
  \bibinfo{person}{Ritesh Agarwal}, \bibinfo{person}{Li Zhang},
  \bibinfo{person}{John Anderson}, \bibinfo{person}{Sarvjeet Singh},
  {et~al\mbox{.}}} \bibinfo{year}{2020}\natexlab{}.
\newblock \showarticletitle{Zero-shot heterogeneous transfer learning from
  recommender systems to cold-start search retrieval}. In
  \bibinfo{booktitle}{\emph{Proceedings of the 29th ACM International
  Conference on Information \& Knowledge Management}}.
  \bibinfo{pages}{2821--2828}.
\newblock


\bibitem[Zhang et~al\mbox{.}(2022)]%
        {zhang2022hot}
\bibfield{author}{\bibinfo{person}{Binjie Zhang}, \bibinfo{person}{Yixiao Ge},
  \bibinfo{person}{Yantao Shen}, \bibinfo{person}{Yu Li}, \bibinfo{person}{Chun
  Yuan}, \bibinfo{person}{Xuyuan Xu}, \bibinfo{person}{Yexin Wang}, {and}
  \bibinfo{person}{Ying Shan}.} \bibinfo{year}{2022}\natexlab{}.
\newblock \showarticletitle{Hot-Refresh Model Upgrades with
  Regression-Alleviating Compatible Training in Image Retrieval}.
\newblock \bibinfo{journal}{\emph{arXiv preprint arXiv:2201.09724}}
  (\bibinfo{year}{2022}).
\newblock


\bibitem[Zhang et~al\mbox{.}(2020)]%
        {zhang2020towards}
\bibfield{author}{\bibinfo{person}{Han Zhang}, \bibinfo{person}{Songlin Wang},
  \bibinfo{person}{Kang Zhang}, \bibinfo{person}{Zhiling Tang},
  \bibinfo{person}{Yunjiang Jiang}, \bibinfo{person}{Yun Xiao},
  \bibinfo{person}{Weipeng Yan}, {and} \bibinfo{person}{Wen-Yun Yang}.}
  \bibinfo{year}{2020}\natexlab{}.
\newblock \showarticletitle{Towards personalized and semantic retrieval: An
  end-to-end solution for e-commerce search via embedding learning}. In
  \bibinfo{booktitle}{\emph{Proceedings of the 43rd International ACM SIGIR
  Conference on Research and Development in Information Retrieval}}.
  \bibinfo{pages}{2407--2416}.
\newblock


\bibitem[Zhang et~al\mbox{.}(2014)]%
        {zhang2014composite}
\bibfield{author}{\bibinfo{person}{Ting Zhang}, \bibinfo{person}{Chao Du},
  {and} \bibinfo{person}{Jingdong Wang}.} \bibinfo{year}{2014}\natexlab{}.
\newblock \showarticletitle{Composite quantization for approximate nearest
  neighbor search}. In \bibinfo{booktitle}{\emph{International Conference on
  Machine Learning}}. PMLR, \bibinfo{pages}{838--846}.
\newblock


\bibitem[Zheng et~al\mbox{.}(2020)]%
        {zheng2020deep}
\bibfield{author}{\bibinfo{person}{Xiangtao Zheng}, \bibinfo{person}{Yichao
  Zhang}, {and} \bibinfo{person}{Xiaoqiang Lu}.}
  \bibinfo{year}{2020}\natexlab{}.
\newblock \showarticletitle{Deep balanced discrete hashing for image
  retrieval}.
\newblock \bibinfo{journal}{\emph{Neurocomputing}}  \bibinfo{volume}{403}
  (\bibinfo{year}{2020}), \bibinfo{pages}{224--236}.
\newblock


\end{thebibliography}

\appendix

\section{Appendix}

In this section, we provide detailed information about retrieval-based applications at Tencent, to facilitate a better understanding of the contributions of the proposed BEBR.

\subsection{Products in Tencent PCG}

\begin{figure}[H]
\centering
\includegraphics[width=0.45\textwidth]{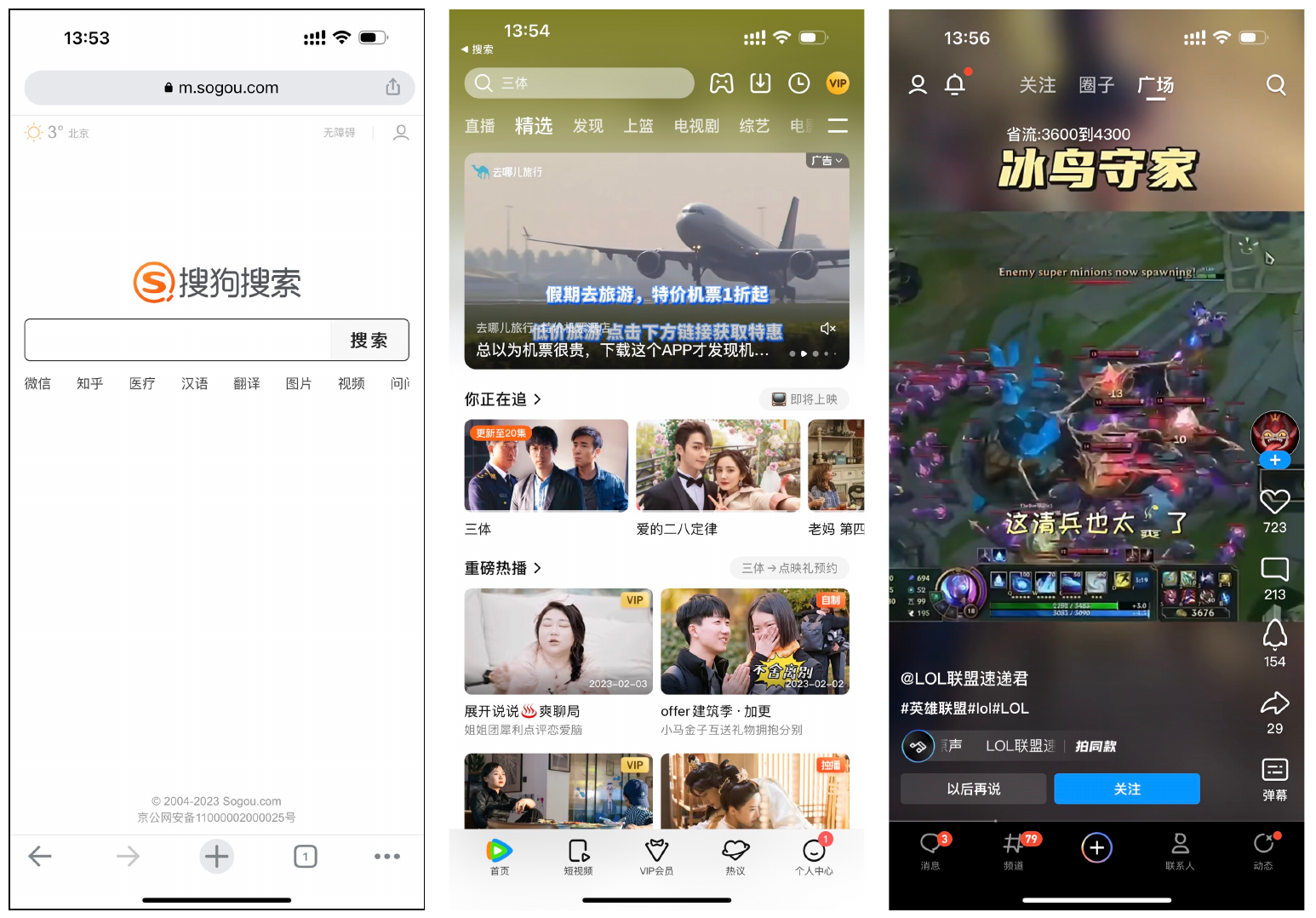} 
\caption{Illustration of partial products at Tencent PCG.}
\label{Fig.products_tencent}
\end{figure}

Tencent is a world-leading internet and technology company that develops innovative products and services to improve the quality of life of people around the world. Our group, Tencent PCG (Platform and Content Group), is a large business group running Tencent's social media, information feed, search, and content platforms, taking Sogou, Tencent Video, and QQ as examples (as demonstrated in Figure \ref{Fig.products_tencent}). 
Large-scale index-based information search (\textit{e.g.}, web/video search) and copyright detection are two fundamental and essential AI capabilities for these products.


\subsection{Information Search}


Almost all products at Tencent require searching for useful information that meets the needs of users. 
For example, the search engine product, Sogou\footnote{\url{http://sogou.com/}}, helps users to acquire information by providing relevant videos, news, and queries under different search sessions. 
Video feeds product returns personalized results of relevant videos to improve user experience. 
Large scale EBR system has been developed as infrastructure at Tencent to provide support to search-related applications. 
However, huge costs in computation and storage as well as search efficiency still remain intractable due to super-scale index and high concurrent queries.


Fortunately, after the deployment of BEBR, up to 73.91\% reduction in both memory and hard disk consumption has been achieved with almost no accuracy loss at the system level, taking one more step towards Green AI. Furthermore, the search efficiency has also been greatly improved with the help of symmetric distance calculation. 
The released resources support further development of emerging functions and sustainable development of the products.

\subsection{Video Copyright Detection}


\begin{figure}[H]
\centering
\includegraphics[width=0.4\textwidth]{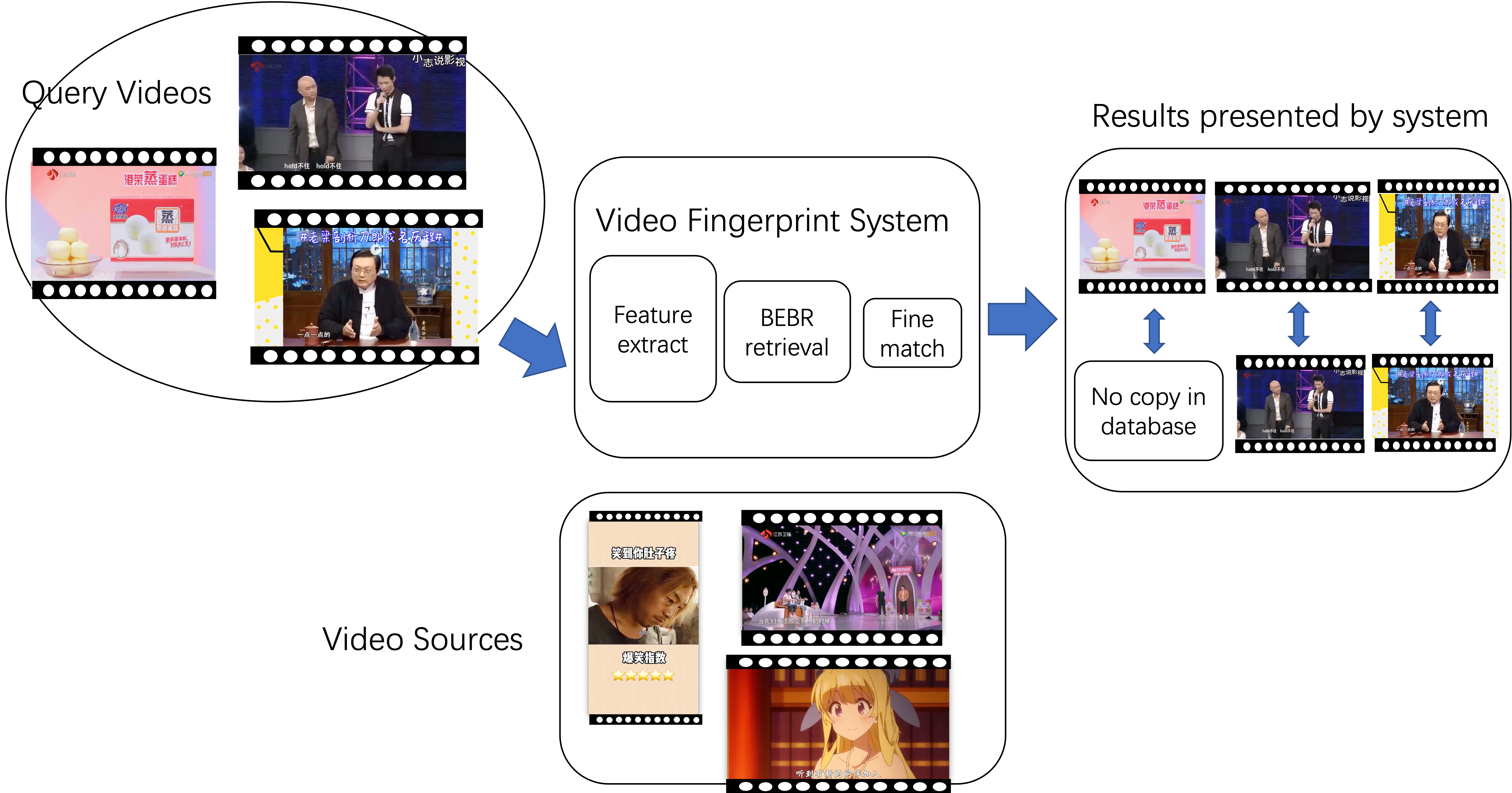} 
\caption{ Illustration of video fingerprint system used in video copyright detection applications. }
\label{Fig.videofingerprint_example}
\end{figure}

Tencent Video (also known as WeTV outside of china) is one of China's largest online video platforms. As of March 2022, Tencent Video has over 1.268 billion monthly mobile active users and 123 million VIP subscribers. Therefore, there are huge traffic videos, and a large number of videos are uploaded on Tencent video every day. This poses great problems for the company in identifying the illicit versions of the original data.

The \textit{video fingerprint} system, which serves copyright detection of Tencent Video, is developed to generate unique and robust identification of the uploaded videos. As shown in Figure \ref{Fig.videofingerprint_example}, the system consists of feature extraction, large-scale retrieval, and fine-grained matching stages, in which the retrieval stage takes almost half of the cost in data computation, memory, and disk storage.
Upon the deployment of BEBR, we considerably reduce the cost of the retrieval stage by around 60\% and the overall cost by around 31\%.
In addition, the fast computation of SDC enables a lower response time, alleviating the pressure on the entire system and improving the user experience.

\end{document}